# Chemical bonding and electronic-structure in MAX phases as viewed by X-ray spectroscopy and density functional theory

Martin Magnuson[1] and Maurizio Mattesini[2,3]

[1]*Thin Film Physics Division, Department of Physics, Chemistry and Biology, IFM, Linköping University, SE-58183 Linköping, Sweden.*
[2]*Departamento de Física de la Tierra, Astronomía y Astrofísica I, Universidad Complutense de Madrid, Madrid, E-28040, Spain.*
[3]*Instituto de Geociencias (CSIC-UCM), Facultad de CC. Físicas, E-28040 Madrid, Spain.*

## Abstract

This is a critical review of MAX-phase carbides and nitrides from an electronic-structure and chemical bonding perspective. This large group of nanolaminated materials is of great scientific and technological interest and exhibit a combination of metallic and ceramic features. These properties are related to the special crystal structure and bonding characteristics with alternating strong M-C bonds in high-density MC slabs, and relatively weak M-A bonds between the slabs. Here, we review the trend and relationship between the chemical bonding, conductivity, elastic and magnetic properties of the MAX phases in comparison to the parent binary MX compounds with the underlying electronic structure probed by polarized X-ray spectroscopy. Spectroscopic studies constitute important tests of the results of *state-of-the-art* electronic structure density functional theory that is extensively discussed and are generally consistent. By replacing the elements on the M, A, or X-sites in the crystal structure, the corresponding changes in the conductivity, elasticity, magnetism and other materials properties makes it possible to tailor the characteristics of this class of materials by controlling the strengths of their chemical bonds.

## Contents

## 1. Introduction

The family of MAX-phase compounds are nanolaminated $M_{n+1}AX_n$ ternary carbides and nitrides, denoted 211, 312, or 413, where n = 1, 2, or 3. The MAX nomenclature is based on the chemical composition of the compounds, were M is an early transition metal, A is a *p*-element that usually belongs to groups IIIA or IVA in the periodic table, and X is C and/or N [1]. These thermodynamically stable nanolaminated carbide and nitride materials possess a remarkable and unusual combination of metallic and ceramic properties [2]. Already in the early 1960s, Hans Nowotny and coworkers in Vienna made X-ray diffraction studies of hot-pressed films and discovered several new carbides and nitrides. Among them, they worked out ternary phase diagrams of Mo-Al-C and found the $Mo_3Al_2C$ phase [3]. These type of phases were named *H-phases* [4], where H denotes *hexagonal* as they were found to have a hexagonal close-packed structure. Shortly afterwards, they also found other H-phases $Nb_3Al_2C$ and $Ta_3Al_2C$ [5] with the same type of crystal structure, as well as $Cr_2AlC$ [6]. In 1967, $Ti_3SiC_2$ and several other 211 and 312 phases were discovered by the group of Nowotny in the form of powder [7] [8] [9]. In the 1960s, Gunnar Hägg in Uppsala formulated stability criteria (Hägg rules) for carbide compounds based on their atomic radii [10]. In the mid 1990s, an unusual combination of metallic and ceramic properties of phase-pure $Ti_3SiC_2$ were reported by Michel Barsoum and El-Raghy [11], who named these structures MAX phases. This led to an enormous increase in scientific interest in these compounds [12]. The first *ab initio* electronic band structure calculation for MAX phases dates back to the work of Ivanovsky *et al.* in 1995 for the $Ti_3SiC_2$ system [13].

Mechanically, the MAX-phases are quite different than their parent binary MX carbides and nitrides. The three different inherent elements in MAX-phases render them more flexible in tailoring properties. The presence of chemically different atomic layers generally increases the strength of the composite material by hindering dislocation motion (i.e., slip plane movement) [1]. The MAX phases, like ceramics, are hard and elastically rigid (much higher stiffness than the parent metals), lightweight, corrosion resistant with high melting points, high strength at high temperature and low expansion coefficient [12]. In addition, MAX phases also exhibit good electrical and thermal conductivity (usually better than the corresponding pure metals), are machinable due to the layered structure, tolerant to thermal shock (~$1400^o$ C), and plastic at high temperatures. Furthermore, MAX phases have low-friction surfaces with high wear resistance. MAX phases are therefore useful in a wide variety of applications. Macroscopically, the combined metallic and ceramic properties are related to the electronic and structural properties of the nanoscale constituent atomic layers. Presently, there are in total more than 70 MAX phases of which more than 50 are known as $M_2AX$ (211) phases, five $M_3AX_2$ (312) phases, and four $M_4AX_3$ (413) phases [14]. MAX phases can be synthesized either as bulk construction materials useful as parts in combustion engines, rockets, and heating elements by sintering at high pressure and temperature or deposited as thin films for surface coatings by physical vapor deposition (sputtering of individual atomic layers). In the latter case, the coatings are used in applications including cutting tools, electrical switches, and diffusion barriers.

Most of the research on MAX phases has incorporated processing and mechanical properties of sintered bulk compounds [1] [12]. However, in many technological applications where, e.g., high melting points, corrosion resistance, electrical and thermal conductivity as well as low-friction properties are required, high-quality thin-film coatings are more useful than bulk materials [14]. The reason for the large interest in MAX phases is the unusual set of combined properties due to the underlying anisotropic structural characteristic and chemical bonding controlling the electronic structure and making the properties tunable by exchanging different elements. Although MAX-phases and related compounds have been studied extensively, detailed understanding of the relationship between electronic structure and physical properties is still lacking. One reason for this lies in the difficulties associated with obtaining accurate electronic structure measurements of internal atomic layers.

**Figure 1:** Elements in the Periodic Table that are known to form $M_{n+1}AX_n$ phases, where M denotes an early transition metal, A is a group A-element and X is either C and/or N.

Here, we review the current understanding of the electronic structure and chemical bonding in MAX phases using laboratory-based X-ray diffraction and synchrotron X-ray spectroscopy compared to state-of-the-art *ab initio* calculations. For understanding the basic physical properties of MAX phases, it is important that phase-pure single crystal materials, e.g., thin films deposited by physical vapor deposition are employed. The goal and topic of this critical review article is an improved understanding and systematization of how the underlying electronic structure and chemical bonding affects the macroscopic properties and how they can be tuned. Explaining the physical properties of MAX phases requires a thorough knowledge of orbital occupation and chemical bonding, as well as the role of phonons [15] [16] and electron correlation effects [17] [18]. By using bulk-sensitive and element-selective X-ray spectroscopy [19] [20], it is possible to differentiate between the occupation of orbitals across and along the laminate basal plane in the interior of the MAX phases.

## 2. Crystal structures and stability of MAX phases

A *nanolaminate* is a material with a laminated - layered - structure in which the thicknesses of individual layers are in the nanometer range. *Inherently* nanolaminated materials have a *crystal structure* that is a nanolaminate as opposed to *artificial* nanolaminates such as thin film superlattices [21]. MAX-phase crystal structures can be described as interleaved atomic layers of high and low electron density. As a consequence of the layered structure, inherently nanolaminated materials often exhibit unique properties, ranging from *mechanical*, *magnetic*, to *thermoelectric* properties. Given their highly anisotropic crystal structure, explaining their physical properties requires an in-depth understanding of the anisotropy and the orbital occupation in the electronic structure.

The layered crystal structures are important for the understanding of the MAX-phase properties. The MAX phases comprise a very large family of materials and presently, there are about 70 different phases, but new ones are still left to be discovered. Most MAX phases have the 211 structure (Ti$_2$AlC, Ti$_2$AlN, V$_2$GeC, etc.), some have 312 structure (Ti$_3$SiC$_2$, Ti$_3$GeC$_2$, Ti$_3$AlC$_2$, Ti$_3$SnC$_2$, Ta$_3$AlC$_2$), and a few have 413 structure (Ti$_4$AlN$_3$, Ti$_4$SiC$_3$, Ti$_4$GeC$_3$, Ta$_4$AlC$_3$, Nb$_4$AlC$_3$, V$_4$AlC$_3$, Ti$_4$GaC$_3$). Figure 1 shows which elements in the Periodic Table can form the MAX phases. The M-elements (early transition metals), include: Sc, Ti, V, Cr, Mn, Zr, Nb, Mo, Lu, Hf, Ta. The A-group (13-16) include: Al, Si, P, S, Ga, Ge, As, Cd, In, Sn, Tl and Ph, while the X-element is either C or N.

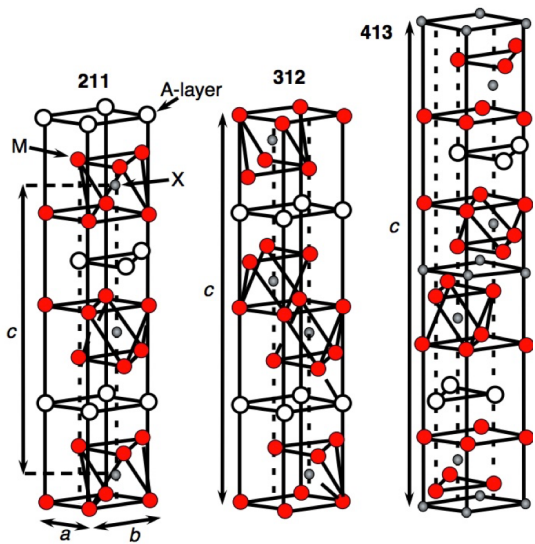

**Figure 1:** MAX-phase unit cell structures of 211 (n=1), 312 (n=2) and 413 (n=3).

In the MAX phases, the hexagonal structure (space group D$^4_{6h}$-P6$_3$/mmc) has two formula units per unit cell. These crystals are characterized by nearly close-packed layers of MX$_6$ octrahedra interleaved with square-planar slabs of atomic layers of A-elements, while the X atoms fill the octahedral sites between the M-atoms. The A-group elements are located at the center of trigonal prisms that are slightly larger than the octahedral sites in order to better accommodate the A-atoms.

Figure shows the crystal structures of the 211 (n=1), 312 (n=2) and 413 (n=3) MAX phases. For the 312 structure, there are two different M sites denoted M$_I$ and M$_{II}$ and for the 413 structure there are also two different X sites, symbolized by X$_I$ and X$_{II}$. Thus, for the 211 structure, there are three inequivalent atoms, while for the 312 structure there are four, and for the 413 structure there are five unique atomic sites. The interleaving pure A-element planes are mirror planes to the zig-zagging M$_{n+1}$X$_n$-slabs. For n=1, ($c \approx 1.3$ nm) there are two M-layers separating each A-layer while for n=2 ($c \approx 1.8$ nm), there are three layers that separate the A-layers. The $c$-axis is much longer than the $a$ and $b$ axis in all three crystal structures. Typically, the $c$-axis is 12-13 Å in 211, 17-18 Å in the 312 structure, and 22-23 Å in the 413 structure.

Table X shows the lattice parameters of thin film MAX-phases in comparison to calculated values with the general gradient approximation (GGA) density functional theory (DFT) using the standard Perdew–Burke–Ernzerhof (PBE) exchange-correlation functional. For Ti$_3$SiC$_2$, the calculated lattice parameters are in good agreement with the lattice constants of bulk materials ($a = 3.068$ Å and $c = 17.669$ Å) [8]. However, in general, the lattice parameters of MAX-phases of single-crystal thin films tend to be slightly shorter than in the case of polycrystalline sintered bulk materials [12]. Due to the difference in repetition of the A-layers, the 211 structure has more metallic and better electrical and thermal conducting properties than the 312 and 413 phases that have more carbide- or nitride-like properties.

Figure shows examples of θ-2θ X-ray diffractograms (XRD) from MAX phase films of $Ti_3AlC_2$, $Ti_3SiC_2$, and $Ti_3GeC_2$ that are used to extract the lattice constants (*a* and *c*) by applying Bragg's law. In all three cases, predominantly $Ti_3AC_2$(000l) reflections are present from the films together with the TiC(lll) reflections [22] and α-$Al_2O_3$(000l) substrate peaks (S) indicating strongly-oriented growth [23]. The observed *c*-axis lattice parameters as presented in Figure correlate well with the reported data of sintered bulk materials [2]. The diffractograms (b) and (c) of $Ti_3SiC_2$ and $Ti_3GeC_2$ also show typical low intensity reflections, marked with arrows, that can be attributed to impurity phases of $Ti_5Si_3C_x$ and $Ti_5Ge_3C_x$, respectively. Diffractogram (a) of $Ti_3AlC_2$ shows a small contribution from $Ti_2AlC$ in the film that is most likely found at the TiC//$Ti_{n+1}AlC_n$ interface since it has also been observed in much thinner films with higher intensity. The low intensity of the impurity peaks compared to the $Ti_3AC_2$ (A=Al, Si, Ge) phase peaks is due to the fact that these impurity concentrations are very small and that their contributions to X-ray spectroscopy measurements can be disregarded. Similar diffractograms were also found for other thin film MAX phases [16] [24] [25] [26] [27].

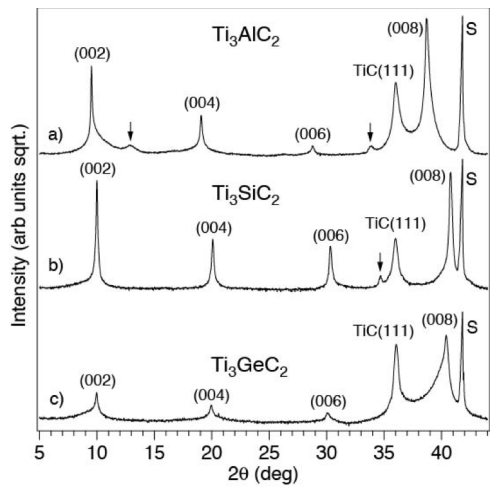

The accuracy of *state-of-the-art* computational methods in predicting the ground-state stability (*e.g.*, formation energies) of MAX phases allow screening of vast numbers of compositional spaces for selected *competing phases*. Recently, a significantly improved tool is available to identify new thermodynamically stable phases and aid experimental synthesis efforts in both bulk [12] and thin-film [14] form. To address the question of which possible phases are expected to occur, quantitative approaches for calculating and predicting the stability of MAX phases are often made by a systematic evaluation of total energies or the *formation energies*. The formation energy is the total energy of the compound minus the sum of the energies of the constituent elements in their stable configurations.

**Figure 2:** XRD diffractograms of the 312 phases $Ti_3AlC_2$, $Ti_3SiC_2$ and $Ti_3GeC_2$ [18].

Another way to probe stability is to calculate *cohesive energies* i.e., the total energy of the compound minus the total energy of the constituent elements at finite separation. A negative energy favors phase formation, while a slightly positive energy likely corresponds to a metastable compound. However, to judge the stability of a hypothetical phase, the calculated energies also have to be compared to all possible competing phases. Furthermore, there are many exceptions, for example, the hypothetically stable $Ti_2SiC$ phase, with the competing phases $Ti_5Si_3$ and $Ti_5Si_3C$, do not occur experimentally. This can be due to a thermodynamic competition from other phases, phonon destabilization by soft modes, or difficulties in identifying all the competing phases. Cover *et al.* [28] studied the stability of 211 phases as well as Dahlqvist *et al* [29] who investigated the stability of 211 phases by using the formation enthalpy of the total energy term in the Gibbs free energy.

Recently, from the calculated formation energies for a very large number of $M_2AX$ phases, general trends in stability have been proposed [30]. Specifically, it was found that phases containing $M$ = Ti, $A$ = group-13 elements, and $X$ = C constitute the largest number of stable $M_2AX$ phases. The primary parameters for empirical design rules were found to be the average electronegativity, differences in ionic radii, differences in ionization potentials for the $A$-site elements, and differences in ionic radii for the $M$-site elements. However, more work is still required in providing design guidelines and future inputs for new stable MAX phases due to competing phases including inverse perovskites and their ternary or quaternary phases.

## 3. X-ray absorption and emission spectroscopy of MAX phases

X-ray absorption spectroscopy (XAS) is a commonly used element-specific technique for probing local and partially unoccupied states of the electronic structure from an excited core-electron. XAS was developed in the 1920s for structural investigations [31]. When probing the pre-edge and near-edge absorption structures, the technique is also referred to as X-ray Absorption Near-Edge Structure (XANES) for solids or Near Edge X-ray Absorption Fine Structure (NEXAFS) for surfaces [32] while the long-range post-edge oscillations are used for structural determinations using Extended X-ray Absorption Fine Structure (EXAFS) [33]. As the XAS technique requires an intense (and often polarized) X-ray beam in a range of photon energies in the vicinity of a core-level energy (*1s*, *2p*, *3p*-shells) that depends on the element of interest, XAS is nowadays performed at synchrotron radiation sources by scanning the photon energy over the absorption edge [34]. The measurements can be performed either in surface-sensitive Total Electron-Yield (TEY) mode or in a more bulk-sensitive Total Fluorescence Yield (TFY) mode [22] at different incidence angles.

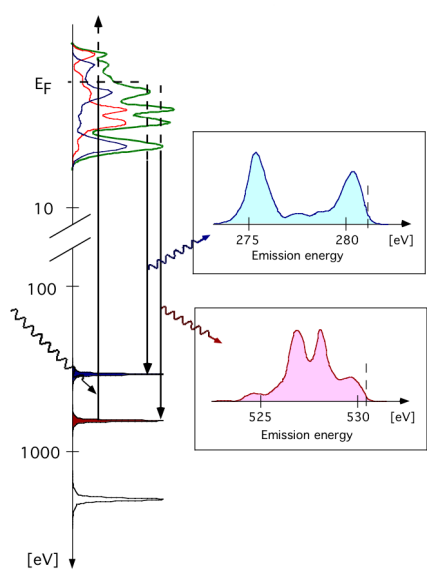

**Figure 4:** Principle of XAS and XES processes.

X-ray Emission Spectroscopy (XES) is an element-specific method to probe the partially occupied electronic structure of materials and was originally developed in the optical wavelength range by Henry Rowland in 1882 by using spherical concave gratings for focusing of light [35]. In the 1920s [36], and 1940s [37], the technique was further developed for determining the valence band structure in the X-ray energy region by measuring the fluorescence emitted when refilling a core-hole that was created in a preceding X-ray absorption process. The use of tunable synchrotron radiation sources has opened up the possibility to study resonant processes in detail. When a core-electron is resonantly excited into a bound state, the XES technique transforms into *Resonant Inelastic X-ray Scattering* (RIXS) [38] [39]. Theoretically, the RIXS process is treated in the Kramers-Heisenberg formula [40] involving transition matrix elements between valence levels and core levels including interference effects between the different states of the core levels.

Figure illustrates the principle of the XAS and XES processes in a material containing two different elements. In the first step of the XES process (photon in, and absorption), corresponding to an XAS process, a core electron is excited or ionized from a selected core-level with a well-defined atomic symmetry. Via the quantum-mechanically allowed *electric-dipole selection rules* ($\Delta l \pm 1$, $p <\!\!-\!\!> d$, $p <\!\!-\!\!> s$) [41], the core-electron is excited to a mixed continuum of unoccupied states in the conduction band or to a certain bound unoccupied state. This is the final step in the XAS process, but simultaneously an *intermediate* step in the XES process. Two different core-levels, (e.g., *1s*, *2p*, or *3p*), corresponding to different elements are indicated in Figure 4 where an electron has been excited from a core-shell with a certian binding energy. In the second step of the XES process (emission, photon out), an electron from either the valence band or a bound conduction state fills the core hole with the simultaneous emission of a photon. The energy of the emitted photon corresponds to the energy difference between the valence band and the core levels which also obey dipole selection rules ($\Delta l \pm 1$, $p <\!\!-\!\!> d$, $p <\!\!-\!\!> s$) to the final state of the XES process. The two different emission paths for filling the core-holes from the mixed states of the valence band to the core-levels yield emitted photons with different energies. By detecting the intensity modulation of the emitted photons over a specific energy window containing an emission line, emission spectra which are characteristic of the elements in the studied material can be obtained. However, for low-Z elements, the fluorescence yield is much lower than the Auger yield and therefore an intense X-ray photon (or electron) beam is needed for excitation of the electrons in the samples. Depending on the energy region of the emission lines, X-ray spectrometers basically have two different designs, either grating-based Rowland-type [35] spectrometers for the soft X-ray region that combines focusing at a concave surface with diffraction in a grating or crystal-based spectrometers for the hard X-ray region.

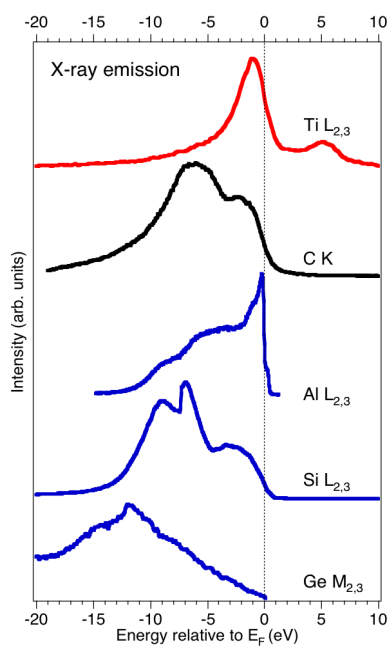

**Figure 5:** Characteristic X-ray emission spectra of the pure elements Ti, C, Al, Si and Ge.

Particularly useful aspects of the XAS technique in TFY mode and the XES/RIXS techniques of great importance for probing buried layers [19] and nanolaminates such as MAX phases containing two or more different elements are the element selectivity and the *large probe and information depth* obtained in fluorescence yield. This makes it possible to probe partial electronic structures from the different elements in the bulk of the materials with negligible contribution from surface contamination if the samples are freshly synthesized.

In 1996, initial XES and XPS measurements on $Ti_3SiC_2$ were performed by Galakhov *et al.* [42] on polycrystalline pressed-powder samples. Similar to other XPS measurements using conventional laboratory sources e.g., by Medvedeva *et al.* in 1998 [43] (see section 4), many early measurements suffered from low energy resolution and poor statistics. Thus, the interpretations of the spectral features were largely hampered by additional broad humps and unknown emission lines from surface oxides and other impurities of, for example, oxygen, as well as additional carbon species at the grain boundaries. In particular, the impurities gave rise to additional intensity close to the Fermi level ($E_F$) as well as very broad features that more resemble characteristic line shapes of amorphous materials than the electronic structure of MAX phases. This was also the case for initial experiments using synchrotron radiation.

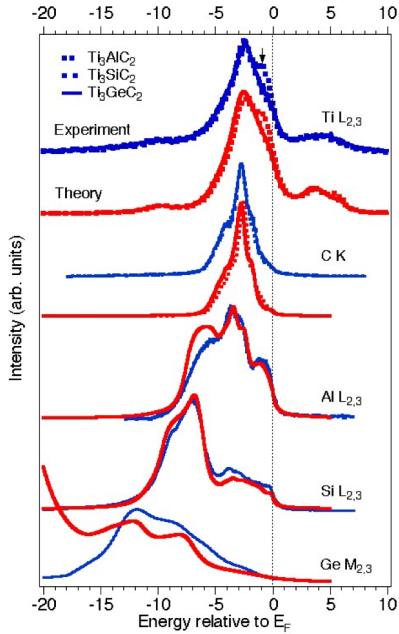

**Figure 6:** Measured Ti $L_{2,3}$, C $K$, Al $L_{2,3}$, Si $L_{2,3}$ and Ge $M_{2,3}$ X-ray emission spectra of $Ti_3SiC_2$, $Ti_3SiC_2$ and $Ti_3GeC_2$ in comparison to calculated spectra [46].

In the early 2000s, the advent of thin-film growth of MAX phases, mainly physical vapor deposition (PVD), but also chemical vapor deposition (CVD), enabled a significant improvement of the crystal structures of the layered MAX phase samples in comparison to sintered bulk samples. Spectroscopic studies of MAX-phases from epitaxially grown samples enabled easier comparison with calculated spectra. In 2005, the first combined XAS and XES measurements using synchrotron radiation were published [23] with a comparison of the electronic structures and chemical bonding of $Ti_3AlC_2$, $Ti_3SiC_2$, and $Ti_3GeC_2$ (see section 4).

Figure shows characteristic XES spectra of the pure elements Ti, C, Si, Al and Ge. As observed, the spectral shapes *differ significantly* depending on the number of valence electrons. For the Ti $L_{2,3}$ XES spectrum, $E_F$ is referenced to the $L_3$ emission line, while the $L_2$ line is observed at higher emission energy. Here, the $L_3/L_2$ branching ratio depends on the $L_2 \rightarrow L_3M$ Coster-Kronig decay changing the initial core-hole population from the statistical 2:1 ratio that is associated with the metallicity of the measured system [44]. For conducting systems, the $L_3/L_2$ ratio is usually significantly higher than the statistical ratio 2:1. Thus, the $L_3/L_2$ branching ratio can be used to compare the metallicity between different materials. Contrary to the Ti $L_{2,3}$ XES spectrum, the C $K$ XES spectrum of pure carbon has a broad spectral shape with a rather broad shoulder, while Si, Al and Ge have more specific peak structures. In particular, crystalline Si exhibits a primary peak feature at 91.5 eV, while metallic Al has a very sharp peak close to $E_F$.

**Table I:** Experimental (calculated) lattice parameters of selected thin-film MAX phases.

| System | Ti$_3$AlC$_2$ | Ti$_3$SiC$_2$ | Ti$_3$GeC$_2$ | Ti$_2$AlC | Ti$_2$AlN | V$_2$GeC | Ti$_4$SiC$_3$ |
|---|---|---|---|---|---|---|---|
| $a$, $b$ [Å] | 3.06 (3.08) | 3.06 (3.08) | 3.06 (3.08) | 3.04 (3.08) | 2.98 (3.01) | 2.99 (3.01) | 3.05 (3.08) |
| $c$ [Å] | 18.59 (18.64) | 17.66 (17.68) | 17.90 (17.84) | 13.59 (13.77) | 13.68 (13.70) | 12.28 (12.18) | 22.67 (22.62) |

Figure shows XES spectra of the 312 MAX-phases $Ti_3AlC_2$, $Ti_3SiC_2$ and $Ti_3GeC_2$ that correspond to the occupied electron bands and can therefore be compared to band-structure calculations including dipole transition matrix elements (density-

functional theory, DFT, Wien2k [45]). As observed, the agreement between experiment and theory is good except for the peak splittings and energy positions in Ge. The Ge $M_{2,3}$ peak splitting is 3.6 eV while the calculated *ab initio* spin-orbit splitting is 4.3 eV (Table IX). Moreover, the calculated shallow Ge *3d* core levels are 3.9 eV closer to $E_F$ and 10 times more intense than in the experiment [46] [23] [26]. The difference can be attributed to screening and relaxation effects. Excitonic effects might also play an important role in determining the peak intensity. Moreover, the Ti $L_{2,3}$ peak splitting is 6.2 eV, while the calculated *ab initio* spin-orbit (so) splitting is 5.8 eV.

These kinds of systematic X-ray spectroscopic studies show that the spectral *shapes* of the internal A-monolayers of Al, Si and Ge in the 312 ternary carbides $Ti_3AlC_2$, $Ti_3SiC_2$ and $Ti_3GeC_2$ [23], are strongly modified by hybridization with neighboring Ti and C atoms in comparison to the corresponding pure elements shown in Figure . However, the energy difference between $Ti_I$ and $Ti_{II}$ is so small that it could not be experimentally resolved, although XES is site-selective. The elemental substitution and corresponding tuning of the valence electron population from the unfilled states of Al to the isoelectronic states in Si and Ge implies that the unusual material properties can be tailored (or "tuned") by the choice of intercalated element.

X-ray spectroscopic methods such as XAS/RIXS and XPS provide experimental values of the weights of the $L_3$ and $L_2$ components and their atomic branching ratios. This serves as important tests of DFT theory [45]. Large deviations in calculated $L_3/L_2$ and $t_{2g}/e_g$ branching ratios beyond the one-electron theory has to be treated as many-body effects including extended exchange and mixed terms between the core states [47]. Generally, the calculated *ab-initio* values of the spin-orbit splitting in band-structure calculations are *underestimated* for the early transition metals (TMs) and overestimated for the late TMs as shown in Table IX. The reason for this is not presently known, but must represent effects beyond the effective one-electron theory in standard DFT e.g., many-body effects.

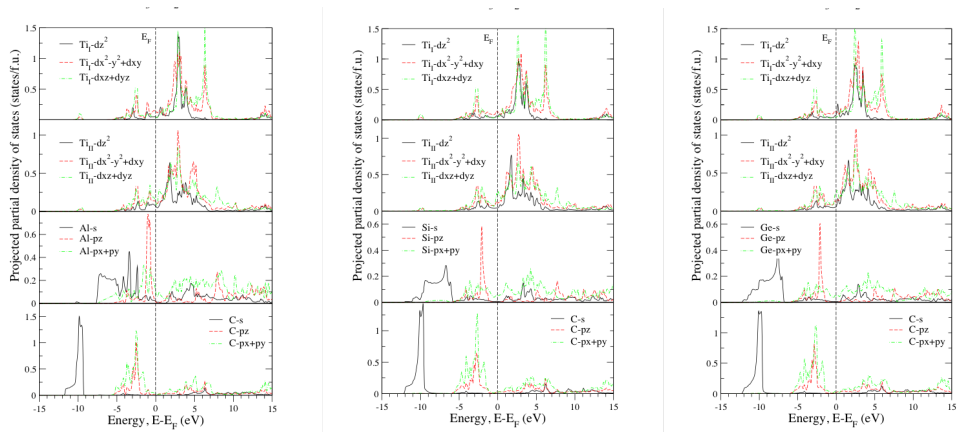

**Figure 3:** Partial angular-momentum projected density of states (pDOS) of $Ti_3AlC_2$, $Ti_3SiC_2$ and $Ti_3GeC_2$.

Theoretically, from a single-particle approach, the branching ratio ($L_3/L_2$) of the $L_3$ and $L_2$ emission lines should be 2:1 if the difference in population and statistical weight of the filled $2p_{3/2}$ (4 electrons) and $2p_{1/2}$ (2 electrons) core shells is considered. However, in XES, the observed $L_3/L_2$ ratio in *3d* transition metals is often significantly higher than the statistical ratio. This is due to the *Coster-Kronig process* named after the physicists Dirk Coster and Ralph Kronig [48]. The Coster-Kronig decay from the $2p_{1/2}$ core-level to the $2p_{3/2}$ level that precedes the X-ray emission process, not only leads to a higher $L_3/L_2$ branching ratio but also to a shorter lifetime and a larger Lorentzian width for the $2p_{1/2}$ core state than for the $2p_{3/2}$ state [44]. The trend in XES branching ratios ($L_3/L_2$ or $M_3/M_2$) in the transition-metal compounds is a signature of the degree of metallicity or ionicity in the systems [22] [49]. A lower branching ratio is thus an indication of higher ionicity (resistivity) in the material. For the MAX phases, a higher branching ratio was observed by Magnuson *et al.* [25] [16] [26] in the basal planes indicating higher metallicity and conductivity than along the *c*-axis.

**Table II:** Measured spin-orbit splittings of common elements in MAX phases.

| so-splitting | XES | XPS | DFT |
|---|---|---|---|
| Sc 2p | 4.8 | 4.9 | 4.64 |
| Ti 2p | 6.2 | 7.4 | 5.77 |
| V 2p | 8.2 | 7.7 | 7.10 |
| Ni 2p | 17.3 | 17.3 | 17.49 |
| Cu 2p | 19.5 | 19.8 | 20.49 |
| Ge 3p | 3.6 | 4.1 | 4.3 |

To achieve a good comparison between theory and experiment, calculated XAS/XES spectra within the one-electron approach often need to be fitted to experimental $L_3/L_2$ or $M_3/M_2$ and $t_{2g}$-$e_g$ branching ratios, as well as experimental spin-orbit splitting values and to an additional broadening for the $L_2$ emission lines due to the Coster-Kronig decay process. A possible solution to this kind of empirical procedure should be tested using multiplet theory [38] [50] or through many-body perturbation theory by solving the Bethe-Salpeter equation (BSE) [51].

## 4. Electronic structure calculations – theory and modeling

A number of comprehensive and systematic theoretical studies of the electronic structure properties have been published for a large number of MAX phases [52] [53] [54] [55] [43]. In 1995, Ivanovsky *et al.* [13] published the first theoretical paper focused on the electronic band structure investigation of $Ti_3SiC_2$ by using the full-potential linear muffin-tin orbital (FP-LMTO) method. Initial theoretical efforts were further concentrated in comparing XPS data together with the computed *first-principles* partial and total density of states (DOS) from different theoretical schemes, such as for example FP-LMTO [12] for $Ti_3SiC_2$ by Medvedeva *et al.* in 1998 [43] and linear combination of atomic orbitals (LCAO) by Zun and Zhou in 1999 [56]. However, with the growing need to understand other known phases and to discovery of new ones, consistent trends in the electro-structural properties were searched within a single computational method (e.g., LCAO [57]). Following the rapid development in condensed matter physics and computational power, more recent works relied on a variety of novel computational methods, such as the full-potential band-structure method (FP-LAPW) [16] [56], pseudo-potential plane-wave method (PP-PW) [58], DFT+U [17] and the hybrid functionals [59] [60]. Thus, not surprisingly, the number of theoretical papers on MAX phases has strongly increased during the last decade. However, it is worth noting that despite the deployment of many different band structure methods, a presently well-accepted agreement has been reached on a number of important conclusions that are valid for all the investigated MAX phases. For instance, a rather strong interaction between *p* and *d*-states of the *M* and *X* atoms has been identified in the region between -2 to -5 eV below the $E_F$. These electronic states originate from the mixing of the *M d* orbitals and the *X 2p* states and give rise to strong directional covalent bonds. Another general characteristic concerns the DOS@$E_F$, which is always dominated by the *d* orbitals of the *M* atoms. At the top of the valence band, in the energy region between -1 to 0 eV from $E_F$, the interaction between the *d*-electrons of the *M* atoms and the *p*-states of the *A* species has generally been computed to be weaker than that between *M* and *X* atoms.

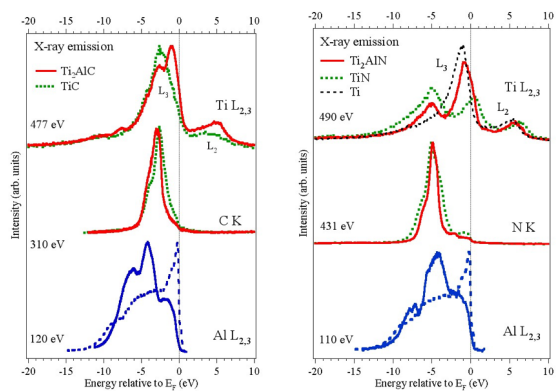

**Figure 4:** XES on $Ti_2AlC$ and $Ti_2AlN$.

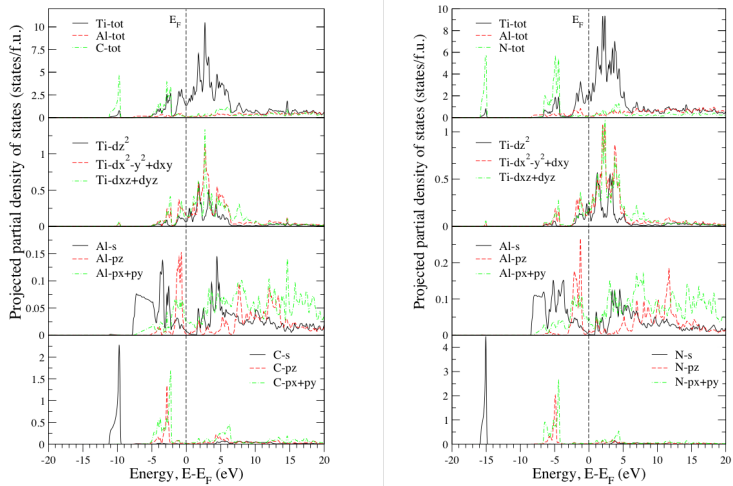

**Figure 5:** Angular-momentum resolved pDOS of $Ti_2AlC$ and $Ti_2AlN$.

Figure 7 shows the angular-momentum-resolved partial density of states (pDOS) of $Ti_3AlC_2$, $Ti_3SiC_2$ and $Ti_3GeC_2$. The peaks between -2.5 eV and -5 eV below the Fermi level ($E_F$) are due to hybridization between Ti *3d* - C *2p_z* orbitals. These overlaps arise from strongly directional covalent bonding. In contrast, the Ti *3d* - Al, Si, Ge *p_z* overlaps are closer to $E_F$ and are relatively weaker in bond strength. The metallic Ti *3d* - Ti *3d* bonding originates from states close to the Fermi level. The Ti *3d* - Si *p_z* bonding appears at lower energy than Ti 3d - Al *p_z* bonding. This results in stronger bonding for Ti *3d* - Si *p_z* than for Ti *3d* - Al *p_z*, thus explaining the reason why $Ti_3SiC_2$ has a higher $C_{44}$ elastic shear modulus than $Ti_3AlC_2$. As we will see in Section 10, the $C_{44}$ elastic modulus reflects the resistance of the crystal to shear in the [010] or [100] plane along the (001) direction, and can thus give important indications about the damage tolerance in solid materials.

A way to judge the quality and predict the power of *ab initio* electronic structure calculations is to compare them with a number of spectroscopic techniques (XPS, XAS, XES, and EELS) that provide important tests of the theoretical results and their accuracy. Most of the experimental XPS, electron energy loss spectroscopy (EELS) and XAS/XES data were often interpreted by using a FP-LAPW [23], FP-LMTO [61] or a multiple-scattering approach [62], providing an overall good agreement between theory and spectroscopic data. In 1998, an XPS study of Kisi *et al.* [63] on a polycrystalline hot-pressed powder sample reported core levels for $Ti_3SiC_2$ having lower binding energies than in the parent carbides TiC and SiC, and this was attributed to the exceptionally screened environment of the high electrical conductivity in Ti₃SiC₂. More recently, Stoltz *et al.* (2003) [64] performed the first XPS experiment using synchrotron radiation on a polycrystalline sample of Ti₃SiC₂. However, these initial XPS experiments largely suffered from oxygen and cadmium impurities as well as additional carbon at the grain boundaries. Although the interpretation of the spectra was hampered by additional broad humps in the data originating from impurity contributions, the experimental valence band spectra were found to be in reasonable good agreement with *state-of-the-art* FP-LMTO band structure calculations of Ahuja *et al.* [61]. Other spectroscopic studies of pressed powder samples were also found to suffer from additional peak structures that made the interpretation unclear.

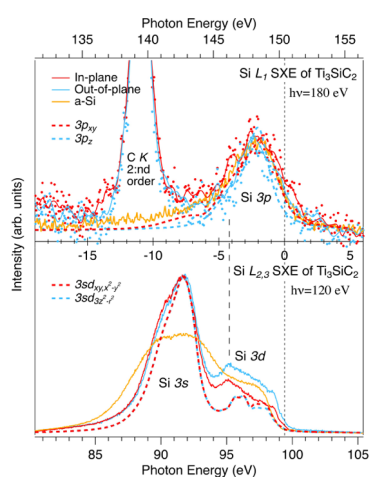

**Figure 6:** Top panel: Si $L_1$ SXE spectra of Ti₃SiC₂ in comparison to pure amorphous Si (a-Si). The dashed curves are corresponding calculated spectra. Bottom panel: Si $L_{2,3}$ SXE spectra in comparison to amorphous Si *[14]*.

Spectroscopic studies of MAX-phases were accelerated when thin film processing with epitaxial growth was introduced that enabled better comparison with calculated spectra. For example, the electronic structure and chemical bonding was investigated using XAS, XES and RIXS using polarized synchrotron radiation. Magnuson *et al.* (2005) [23] investigated, both experimentally and theoretically, the spectral shapes of the *A*-layer (Al, Si, and Ge) in the 321 ternary carbides Ti₃AlC₂, Ti₃SiC₂, and Ti₃GeC₂ and identified a strong hybridization with the neighboring Ti atoms in comparison to the corresponding pure elements. Using the same kind of cross-interdisciplinary methodology (experiments and theory), Ti₂AlC was also compared to TiC [22]. In agreement with *ab initio* calculations, the M 3*d*-C 2*p* and M 3*d*-C 2*s* bonding regions were found to be lower in energy and therefore stronger than in TiC. A very similar situation was also disclosed for Ti₄SiC₃ [65]. When comparing Ti₂AlC, Ti₂AlN, TiN [24], and AlN [66] the electronic structure and chemical bonding were found to be considerably different [67] [68]. Nitrides have deeper bond regions and therefore stronger bonds.

EELS spectroscopy can also be used to study the electronic structure of MAX phases. In 2005, Hug *et al.* [69] used EELS to investigate Ti₂AlC, Ti₂AlN, Nb₂AlC, and TiNbAlC and obtained good agreement with FP-LAPW and full multiple scattering theoretical calculations. Mauchamp *et al.* [70] used EELS to study Ti₂AlN and successfully probed the anisotropy in its dielectric response. Once again, the experimental results were in good agreement with *ab initio* calculations based on DFT. Nonetheless, theoretical modeling is not always an easy task, and very often reasonable results are only achieved if one goes beyond the simple ground state single-particle approach. For instance, accounting for core-hole and phonon effects in XAS, XES and RIXS requires extended unit cells (i.e., supercells) including valence to core-level transition matrix elements with Clebsch-Gordan coefficients. For the computations of the XES spectra, however, the *final-state* rule can be applied, [71] where no core-hole is created at the photoexcited atom. Phonon vibrations [16] and excitonic effects [51] must also be considered when identifying weights of different spectral components and atomic branching ratios in experimental data providing important tests of theory [45]. This fact and the large deviation in calculated $L_3/L_2$ and $t_{2g}/e_g$ branching ratios beyond one-electron theory has to be treated as many-body effects including extended exchange and mixed terms between the core states [47]. A related issue is the large discrepancy between theory and experiment for the energy positions and intensities of the *shallow 3d core levels* in Ga [72] and Ge [26] [25], where additional on-site Coulomb interaction is needed to obtain physical agreement. A similar problem has been found in the Cr-containing MAX-phases, such as $Cr_2GeC$, where the magnetic Cr *d*-states must be carefully handled either within an *ad hoc* +U potential [17] or under a hybrid functional formalism [73].

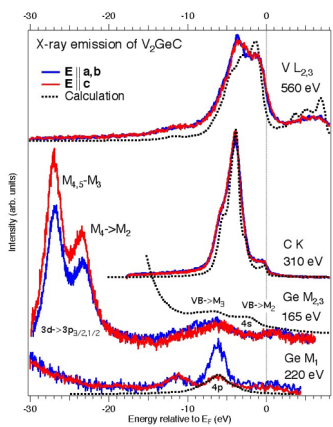

**Figure 7:** V $L_{2,3}$, C $K$ and Ge $M_{2,3}$ XES spectra of $V_2GeC$.

In addition to the importance in addressing the spectroscopic data, the theoretical knowledge of the partial and total DOS found considerable application in assessing the intrinsic stability of a crystal. For a metallic MAX phase, the topological local features of $DOS@E_F$ represent a key quantity to address phase stability. Specifically, a local minimum at $E_F$ signifies higher structural stability, while a local maximum at $E_F$ implies structural instability. This is because a local DOS minimum at $E_F$ acts as an energy barrier for electrons below the Fermi level to move into the empty states of the conduction bands. Such qualitative criteria explain in part why it is difficult to synthesize pure $Ti_2SC$ and $Ti_2PC$ [57] [74] [75]. In general, the intrinsic structural instability correlates well with the simple mechanism of valence electron fillings. As formerly observed by Hug [74], most of the MAX phases have a $DOS@E_F$ that increases with the filling of valence electrons from *A*, *M* and *X* elements. $Ti_2SC$ represents an exception to this behavior, as its computed $DOS@E_F$ is smaller than $Ti_2PC$, even though sulfur has one more valence electron than phosphorous [74]. Nonetheless, we here remind that the above criterion does not go further than a mere qualitative index and other factors, such as the competition from other phases in the same phase diagram (*e.g.*, $Ti_2SiC$), must be taken into account to determine the real stability of a crystal [14].

The mechanical properties of MAX phases (Section 10) are controlled by complex mechanisms of deformation, which are intimately related to the features of the electronic structure and chemical bonding. Thus, theoretical modeling also represents an important tool to investigate and probe the electron-mechanical correlation in MAX phases. As discussed in detail in Section 6, marked differences in the computed charge density distribution are found for the MAX phases, which point to a strongly anisotropic chemical bonding environment. In the case of $Ti_3SiC_2$ crystals, plasticity is also unusually anisotropic, and this characteristic can be traced back to its intrinsic chemical bonding anisotropy [12] [76]. The study of plasticity is of great importance to understand the formation of dislocation loops and the nature of brittle-ductile and brittle-fracture transitions [14] [77]. In this sense, *ab initio* calculations were carried out to model the brittle fracture in $Ti_3SiC_2$, so as to put forward its microscopic mechanism [78] [79] [80]. Also, in order to comprehend the oxidation and corrosion properties of $Ti_3SiC_2$, computations of cleavage energies from different models and methods (Spanish Initiative for Electronic Simulations with Thousands of Atoms (SIESTA) code [81], Vienna Ab-initio Simulation Package (VASP) code [79], and FP-LAPW code [80]) have led to very similar results, showing that Ti-C bonds are twice as strong as those of Ti-S. All these *ab initio* calculations [12] [76] [14] agree in that a crack in the $Ti_3SiC_2$ crystal will likely originate between $Ti_{II}$ and Si layers where the cleavage energy is two times lower than any other atomic layer.

The chemical bonding schemes in carbides and nitrides is quite different, and this can be rapidly appreciated by looking at their electronic state distributions (Fig. 9). Starting with the binary systems, TiC and TiN, the distance to the Ti-X peak from $E_F$ is two times larger in TiN at -5 eV in comparison with TiC at -2.6 eV (see the balanced crystal overlap population (BCOOP) analysis in Section 7). Thus, the covalent Ti *3d*-N *2p* bonding in TiN is significantly stronger than the Ti *3d*-C *2p* bonding in TiC. By further analyzing the partial DOS for $Ti_2AlC$ and $Ti_2AlN$ (Fig. 9), their crystal overlap population data (Section 7) and bond lengths (Section 6), the same difference as in the binaries is observed for $Ti_2AlC$ [27] and $Ti_2AlN$ [24]. This general difference tendency is also confirmed by the energy shift in the experimental X-ray emission spectra of $Ti_2AlC$ in comparison to $Ti_2AlN$ (Figure ).

## 5. Anisotropy and polarization dependence

The electronic structure *anisotropy* in $V_2GeC$ was studied [21] with the complimentary XES+DFT methodology, demonstrating the spectral anisotropy of the different in-plane and out-of-plane bonding orbitals. The polarization dependent Ge $M_1$ and Ge $M_{2,3}$ edges were found to be very sensitive to the symmetry and anisotropy of the V atom coordination shell, both in- and out-of-plane.

Figure 10 shows V $L_{2,3}$, C $K$, Ge $M_1$ and $M_{2,3}$ XES spectra of $V_2GeC$, [26] measured using linearly-polarized synchrotron radiation both in the basal *ab*-plane and along the *c*-axis. The V $L_{2,3}$ spectra (top) were found to be sensitive to the local coordination of the V atoms in- and out-of-plane. The V $L_{2,3}$ - C *2p* hybridization region at -4 eV is deeper than for $Ti_2AlC$ (-2.6 eV) as shown in Figure . As seen in other 3d containing MAX phases, the conductivity depends on the V *3d* states at $E_F$.

Contrary to the V $L_{2,3}$ and C $K$ XES, the anisotropy of the Ge $M_1$ and $M_{2,3}$ XES spectra was found to be large. As illustrated at the bottom of Figure 7, the difference between the in-plane Ge $4p_{xy}(\sigma)$ and out-of-plane $4p_z(\pi)$ bonding orbitals probed by the Ge *4p* -> *3s* and *3d* -> *3p* transitions was substantial, while the Ge *4s* -> *3p* transitions were basically isotropic. [26]

Figure 11 shows an interesting angular-dependent XAS and RIXS study on phase-pure thin-film $Ti_3SiC_2$ where it was shown to be possible to probe the electronic states at grazing and near-normal incidence angles and differentiate the out-of-plane $p_z$ and $d_{3z2}$ states from the in-plane $p_{xy}$, $p_{xy}$, $d_{x2-y2}$ states [14]. Then, XES measurements were made at $15^o$ and $75^o$ incidence angles using linearly-polarized X-rays from a synchrotron. For the Si $L_1$ XES (Top panel, Figure), the anisotropy in the weak Si *3p* states is shown by the difference between the three $3p_{xy}$-σ orbitals that are spread out between 0 and -5 eV below $E_F$ and, the single $3p_z$-π orbital that is more localized around -2 eV below $E_F$. The bottom panel shows the Si $L_{2,3}$ XES with isotropic *3s* states that have a peak at -7.5 eV. Note that the Si $L_{2,3}$ spectra of $Ti_3SiC_2$ has very different shape than single-crystal bulk, as well as amorphous Si. In particular, the Si *3d* states exhibit significant anisotropy with 73% larger intensity along the *c*-axis at $E_F$. However, this is not reproduced in ground-state DFT calculations at 0 K without taking phonons into account. As will be shown in Section 9, the calculated phonon frequency spectra (PhDOS) of the Si atoms in the *ab*-basal in-plane (Si-x, Si-y) phonons have 3-4 times lower frequency (3.3 THz) than the out-of-plane phonons (10-12 THz) along the *c*-axis (Si-z). In fact, the Si-atoms are known to act as "rattlers" and the Si *3d* XES character weighting should be compared to the partial DOS when the core-excited atoms are displaced. Theoretically, a substantial anisotropy within 1 eV from $E_F$ is found when the Si atoms are moved along the *c*-axis with the *static displacement method*. On the contrary, displacement along *ab*-basal plane gives negligible anisotropy.

Generally, the anisotropy in the electronic structure is important for understanding the origin of the negligible Seebeck coefficient in $Ti_3SiC_2$. In fact, the Seebeck coefficient *S* (*i.e.*, a measure of the thermoelectric property or thermopower) in nanolaminated $Ti_3SiC_2$ crystals can be traced to anisotropies in element-specific electronic states. A larger number of in-plane states at $E_F$ is associated with a positive contribution to the Seebeck coefficient in the basal *ab*-plane, while there is a negative contribution to *S* by out-of-plane states. The opposite signs are related to electron- and hole-like bands near $E_F$ and the average contribution to *S* is zero in $Ti_3SiC_2$. These results provide experimental evidence explaining why the average Seebeck coefficient of $Ti_3SiC_2$ in polycrystals is negligible over a wide temperature range (see Section 8).

**Table III:** Calculated bond lengths in a few selected MAX-phases in comparison to binary compounds.

| Bond type | $M_I$ - X [Å] | $M_{II}$ - X [Å] | $M_{II}$ - A [Å] | A - X [Å] |
|---|---|---|---|---|
| TiN | 2.129 | | | |
| TiC | 2.164 | | | |
| VC | 2.082 | | | |
| $Ti_2AlN$ | | 2.088 | 2.834 | 3.826 |
| $Ti_2AlC$ | | 2.117 | 2.901 | 3.875 |
| $V_2GeC$ | | 2.040 | 2.634 | 3.506 |
| $Ti_3AlC_2$ | 2.201 | 2.086 | 2.885 | 3.802 |
| $Ti_3SiC_2$ | 2.189 | 2.097 | 2.694 | 3.608 |
| $Ti_3GeC_2$ | 2.212 | 2.108 | 2.763 | 3.698 |

Recently, another interesting study was made on $Cr_2GeC$ using a combination of XAS and XES [25] [17]. A significant anisotropy was observed in both XAS and XES of Cr with larger intensities in the basal *ab*-plane than along the *c*-axis at $E_F$. More Cr *3d* states were observed in-plane than along the *c*-axis with much more empty hole-like C *2p* states in the basal *ab*-plane than along the *c*-axis around $E_F$. Anisotropy was also observed for the Ge *4p* states in $M_I$ XES with three $4p_{xy}$-σ orbitals spread out between 0 and -5 eV below $E_F$, while a single $4p_z$-π orbital more localized around -2.5 eV below $E_F$ was observed along the *c*-axis. A particularly interesting feature is the isotropic *4s* states observed at -12.5 eV in the Ge $M_{2,3}$ XES data. The Ge *4s* states exhibit significant intensity that is not reproduced in ground-state DFT calculations at 0 K. Generally, the *4s/3d* intensity ratio of Ge and Ga is not in agreement between experiment and DFT calculations [23] [26] [72]. A complicating factor may be the strong electron-phonon coupling with a Ge oscillation along the *c*-axis that has a higher frequency than along the *ab*-plane (see Section 9). However, this effect cannot account for the large difference between experiment and theory in determining the number of states at $E_F$. Experimentally, $Cr_2GeC$ has 22 states at $E_F$, while the DFT calculations exhibit only 7.7 states. Instead, the large intensity at $E_F$ in $Cr_2GeC$ is related to an intensity redistribution from Ge *3d* to *4s* states. The greater intensity of the Ge *4s* states observed experimentally explains the large difference between experimental and calculated DOS at $E_F$.

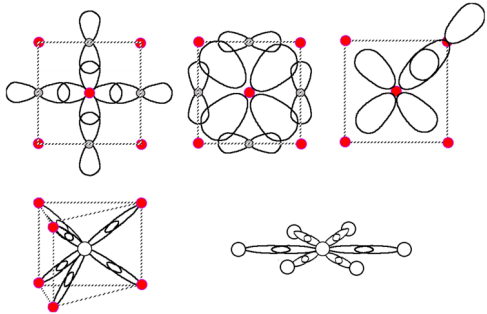

**Figure 8:** Illustration of covalent bonding orbitals in MX (top row). The $pd_\sigma$ bonds are overlaps between the M $e_g$ and the X *2p* orbitals. The $pd_\pi$ bonds are overlaps between the M $t_{2g}$ and the X *2p* orbitals. The $dd_\sigma$ bonds are overlaps between the M-M $t_{2g}$ orbitals. Bottom: $p_z$ and $p_{xy}$ orbitals at the A-atoms. The A-atoms are located in the center of trigonal prisms surrounded by M-atoms.

While the quantitative agreement between experimental and calculated spectra is not always perfect, as a general tendency, the XAS and XES spectral shapes are mostly consistent with DFT predictions. Due to the very large structural difference between the *c*-axis *versus* *a-b* axis, an anisotropy in the electronic structure is expected that should affect the transport properties as well. Mauchamp *et al.* investigated the anisotropy in the resistivity of $Ti_2AlC$ [82] and $Ti_2AlN$ [70] using EELS in comparison to DFT calculations and found a relatively strong anisotropy.

Furthermore, Mattesini and Magnuson showed that $Cr_2GeC$ has clear carrier-type anisotropy [17] i.e., that hole-type carriers are responsible for the transport properties within the basal *ab*-plane, while along *c*-axis the electrons are the dominant charge carriers. In addition, the positive Seebeck coefficient of $Cr_2GeC$ [25] suggests that *p*-type carriers along the *ab*-plane direction provide the main contribution for the bulk Seebeck coefficient of $Cr_2GeC$ and related systems as further discussed in section 8.

## 6. Chemical bonds in MAX-phases

As in the case of the binary MX compounds, chemical bonding in the MAX-phases is a combination of metallic, covalent, and ionic bonding. M-X bonding is strong in the MAX phases as for the binary MX compounds, but the M-A bonds are weaker than the M-X bonds and the density of states at the Fermi level is dominated by the *3d* states of the M-atoms. Figure shows the different types of covalent bonding orbitals in MX compounds. As observed in the XAS spectra at the $2p_{3/2,1/2}$ thresholds of Ti, the crystal-field splits the anti-bonding Ti *3d* band states into the $t_{2g}$ and $e_g$ symmetries, where the peak structures are separated by ~1.5-1.8 eV, depending on the compound. As illustrated in Figure , the lobes of the M-$e_g$ orbitals extend toward the neighboring X atoms (C or N) and form $pd_\sigma$ bonds with the *2p* orbitals of the neighboring X atoms. The lobes of the M $t_{2g}$ orbitals form $pd_\pi$ bonds with the overlap of the *2p* orbitals of the adjacent C atoms. These covalent bonds exhibit the dominate bonding contribution. The same lobes also form metal-metal $dd_\sigma$ bonds with the $t_{2g}$ orbitals of the adjacent M atoms. The metallic bonding occurs to a greater extent at anti-bonding energies above the $E_F$ and is therefore relatively weak. Table VII compares the calculated bond lengths in a few common selected MAX-phases with the binary MX compounds TiC, TiN and VC. Although the bond lengths are similar, there are important differences in the trends. For example, the $M_{II}$-X bonds are shorter than in the binary MX compounds.

Similar to TiC, the chemical bonding in the MAX phases have a mixed covalent-ionic-metallic nature which result in the combination of ceramic and metallic properties. It is well-known that ceramics are usually characterized by covalent and/or ionic bonds which result in macroscopic properties as, for example, low electrical conductivity, high melting point, brittleness and high hardness. By stacking the A-element (e.g., Al, Si, Ge) in a sequence A-Ti$_{II}$-C-Ti$_{II}$-C-Ti$_{II}$-A in the TiC matrix, the electrical conductivity and other properties useful in, for example, high-temperature applications are greatly improved. The high melting temperature is associated with the strong covalent and ionic bonds, while the electrical and thermal conductivity are due to the strong metallic bonds. The Ti-Ti bonds of the type Ti$_{I}$ and Ti$_{II}$ atoms in the 312 and 413 structures are quite different and play different roles for the electrical conductivity due to the metallic bonds. This is because of the fact that the Ti$_{II}$ atoms are bonded to both C and A elements, while Ti$_{I}$ atoms are only bonded to C. Similar to TiC, the covalent Ti$_{I,II}$-C bonding is very strong while the Ti$_{II}$-A bonding is much weaker. The A elements also form covalent bonds with each other. In single-crystals, the metallic conductivity properties are anisotropic and depend on which crystal direction they are measured.

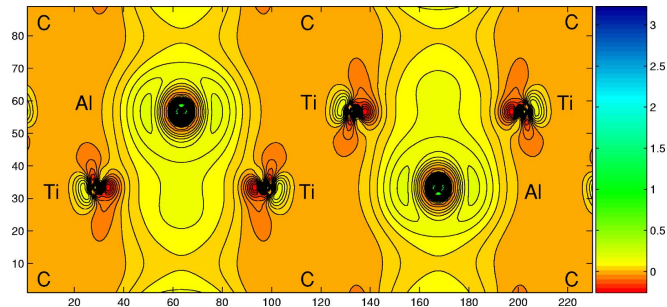

**Figure 13:** Computed electron density difference plot along the [110] plane between Ti$_2$AlC and Ti$_2$C$_2$ (TiC) in the same crystal geometry *[27]*. The difference density plot was obtained by subtracting the charge densities in the 110 diagonal plane of the hexagonal unit cell. Positive values (light green) mean gain of density and negative values (dark red) loss of density (e/Å$^3$).

Generally, common bond characteristics can be identified in MAX-phases. Just as in TiN and TiC, in MAX-phases there are covalent bonds, metal bonds and ionic bonds that depends on the difference in electronegativity between the elements involved. The M *3d* - X *2p* bonds are much stronger than the M *3d* - A *p* bonds, and more charge is found in the M - A bond when the A-element is Si or Ge, instead of Al. An increasing amount of charge is observed in the M - X and M — A bonds when X is N instead of C. A general tendency for these kind of nanolaminated materials is that a weaker bond in one direction from M tends to be compensated by a stronger bond in another direction. It is generally found that the M - X bonding is stronger in the MAX-phases than in their parent binary TiN and TiC phases. These characteristics make the MAX phases a special class of materials, whose physical properties can be tuned *via* bond engineering. By changing the A and X-elements, it is possible to modify the bond strengths and tailor these materials for desired macroscopic properties.

An interesting way to obtain straightforward insights into the chemical bonding nature of MAX phases is to make use of the *ab initio* computed electron density maps. Of special interest for engineering the bonding type and strength in MAX phases is the study of *electron density difference plots*, which provide useful information on the local chemical bonding relative to the parent binary systems. An illustrative example is given in Figure, where the calculated electron density difference plot between $Ti_2AlC$ and $Ti_2C_2$ is shown. In this case, $Ti_2C_2$ denotes a reference binary phase (*i.e.*, TiC) where the Al has been replaced by C in the same 211 crystal structure. When introducing Al atoms into the $Ti_2C_2$ crystal structure, an anisotropic charge variation around the Ti atoms is observed in Figure. In particular, in the direction along the Ti-Al bond (45° angle to the corners of the plot), an electron density withdrawal is noticed (see the red/dark area around Ti) from Ti to Al that indicates the formation of the Ti-Al bonds. The consequence of this electronic movement is the creation of polarization in the neighboring Ti-Ti bonding which reduces its strength. The insertion of the Al atoms in the $Ti_2C_2$ structure introduce a local anisotropic electron density distribution around the Ti atoms resulting in charge-modulation along the Ti-Al-Ti-Ti-Al-Ti zigzag bonding direction that propagates throughout the entire unit cell. The charge transfer from Ti toward Al, which is an indication of an ionic contribution to the bonding, is in agreement with the measured XPS core-level shifts [27] and the BCOOP analysis discussed in Section 7. Finally, the charge-density difference is zero at the carbon atoms located at the corners of the density plot in Figure. This suggests that C atoms do not respond significantly to the introduction of Al planes, and implies that Al substitution only results in local modifications to the charge density, and possibly a weak Al-C interaction. A very weak Al-C bond has also been observed experimentally [83].

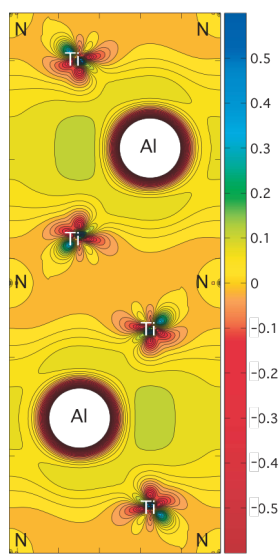

**Figure 14:** Calculated electron density difference plot between $Ti_2AlN$ and $Ti_2N_2$ (TiN) in the same crystal geometry. Positive values (light green) mean gain of density and negative values (dark red) loss of density ($e/Å^3$). The plot was obtained by subtracting the charge densities in the diagonal plane of the hexagonal unit cell.

The same kind of methodology has been applied to $Ti_2AlN$ [24]. Figure 14 shows the computed electron density difference plot between $Ti_2AlN$ and $Ti_2N_2$, where in the latter case N has replaced Al in the same 211-crystal structure. When introducing the Al atoms into the $Ti_2N_2$ matrix, an electron density loss is observed at the Al atomic sites. Around the Ti atoms, an anisotropic charge density variation is obtained with a considerable loss of electron density. In contrast, electron density gain in the direction toward the N and Al atoms is observed indicating the formation of Ti-N and Ti-Al bonds. As in the case of $Ti_2AlC$, the consequence of this electron density flow (i. e., charge transfer) is the creation of polarization with a loss of electron density on the neighboring Ti-Ti bonding, thus reducing its strength. The introduced local anisotropy in the electron density distribution around the Ti atoms generates a charge modulation along the Ti-Al-Ti zigzag bonding direction. The yellow-light green areas around the N atoms in Fig. 14 imply a gain of electron density primarily from Ti, but also from Al. This shows that the N atoms respond significantly to the introduction of Al planes, implying that Al substitution for N results in local modifications of the charge density pattern. Note that in comparison with C in $Ti_2AlC$, N in $Ti_2AlN$ is more electronegative and withdraws a larger fraction of electronic density from Al, leading to a stronger Al-N interaction. The charge transfer from Ti and Al toward N is in agreement with the BCOOP analysis presented in Section 7.

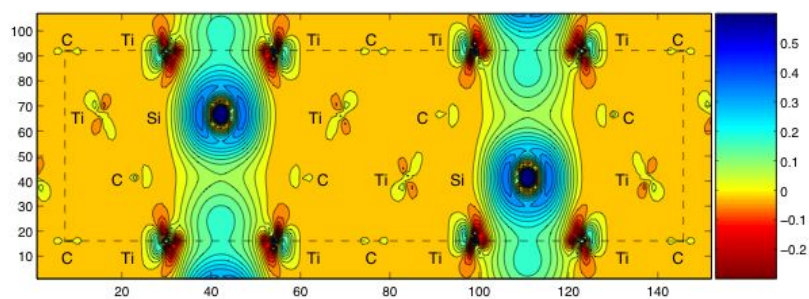

**Figure 15:** Calculated charge-density difference between $Ti_4SiC_3$ and $Ti_4C_4$ (TiC) in the same crystal geometry [65]. A carbon atom is located at each corner of the plot where the charge-density difference is zero. The difference density plot was obtained by subtracting the charge densities in the [110] diagonal plane of the hexagonal unit cell.

Another illustrative example of electron density analysis is given in the paper of Magnuson *et al.*, in 2006 for a 413 MAX phase [65]. Figure 15 shows the calculated electron density difference between $Ti_4SiC_3$ and $Ti_4C_4$, where in the latter system C has substituted for Si in the $Ti_4SiC_3$ crystal. Again, the introduction of Si atoms into the $Ti_4C_4$ phase creates an anisotropic charge variation around the Ti atoms that are close to Si. A close look at Fig. 15, in the direction of the Ti-Si bond, shows that an electron density withdrawal from Ti to Si is observed (see the dark red area around Ti atoms), revealing the formation of Ti-Si bonds. The effect of such an electron-density displacement is the polarization of the neighboring Ti-Ti bonding. Thus, the insertion of Si atoms into the $Ti_4C_4$ structure introduces an anisotropic electron density distribution primarily in a thin sheet containing Ti and Si atoms, resulting in an overall charge modulation along the Ti-Si-Ti zigzag bonding direction of the unit cell. It is also observed that the computed charge-density difference vanishes at the C atoms, revealing that C atoms respond very little to the inclusion of the Si planes. This means that Si substitution only results in local adjustments to the charge density, and probably to a weak Si-C interaction.

# 7. Balanced Crystal Orbital Overlap Population analyses

A useful concept in DFT is to simulate the strengths of chemical bonds by applying BCOOP analysis using the full-potential linear muffin-tin orbital method [84]. This makes it possible to compare the strength of two similar chemical bonds by comparing the *integrated areas* under the BCOOP curves and peak positions from the Fermi level.

In order to understand the nature of the chemical bonds in binary MC compounds and related MAX-phases, the BCOOP between the M *3d* and C *2p* orbitals can be calculated as illustrated in Figure 16 for $Ti_3AlC_2$, $Ti_3SiC_2$ and $Ti_3GeC_2$ in an analysis presented by Magnuson *et al.* in 2005 [23]. For the MAX-phases, the overlap of the *3d* orbitals of the $Ti_I$ and $Ti_{II}$ with the C *2p* orbitals as well as the overlap of the $Ti_{II}$ 3d orbitals with the A-elements (Si, Al, Ge) were calculated. BCOOP is positive for bonding states and negative for anti-bonding states as observed in Figure [23].

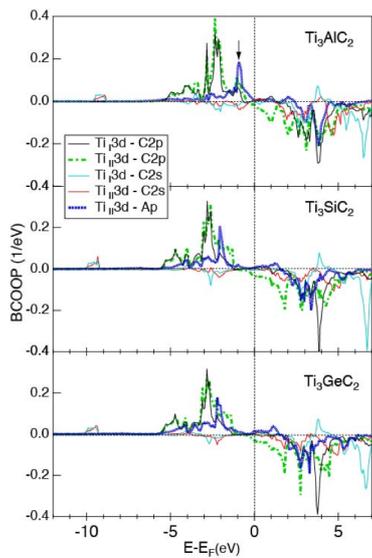

**Figure 16:** Calculated BCOOP plots for Ti₃AlC₂, Ti₃SiC₂ and Ti₃GeC₂.

By comparing the BCOOP curves for TiC with those for the corresponding MAX-phases, it is possible to identify the relative strengths of similar chemical bonds. The *intensities* of the peaks of the BCOOP curves, and their *areas*, provide an indication of the concentration and the amount of overlap between orbitals. In addition, the peak energies give an indication of the strength of the bonding. Thus, by determining the areas under the BCOOP curves and the positions of the peaks in Figure 16, it can be concluded that the Ti $3d$ - C $2p$ overlap is much stronger and more pronounced below $E_F$ than the Ti $3d$ - $A$ (Al, Si, Ge) overlap which has a weaker character. The comparison of the BCOOP curves for the different systems shows that the $Ti_{I,II}$-C BCOOP curves of Ti₃AlC₂ are the most intense and are somewhat shifted toward $E_F$. In general, the areas under the $Ti_{II}$-C peaks are larger than for the $Ti_I$-C peaks, indicating a stronger bond. An interesting observation is that the $Ti_{II}$-Al BCOOP peak is located at about 1 eV below $E_F$, while the $Ti_{II}$-Si and $Ti_{II}$-Ge BCOOP peaks are located approximately 2 eV below the $E_F$. This indicates that the $Ti_{II}$-Al bond is weaker than the $Ti_{II}$-Si and $Ti_{II}$-Ge bonds, which is confirmed by the differences in bond lengths. However, the bond lengths of $Ti_I$-C in Ti₃AC₂ (A with Si, Al, Ge) are longer than those of $Ti_{II}$-C for all the MAX-phases. Since the bonding environments of $Ti_I$ and $Ti_{II}$ atoms are quite different, the stronger bonds associated with the $Ti_{II}$ atoms is not surprising.

Differences in the chemical bonding between Ti₂AlC, Ti₃AlC₂ and TiC structures have also been investigated [27]. From the computed BCOOP plots in Fig. 17, it is possible to compare the strength of two similar chemical bonding types. Observing the areas under the BCOOP curves and the distances of the main peaks from $E_F$, the Ti $3d$–C $2p$ bond is much stronger than the Ti $3d$ – Al $3p$ bond in both Ti₂AlC and Ti₃AlC₂. Hence, Ti atoms bond more strongly to C than Al, which gives rise to a stronger Ti-C bond for $Ti_{II}$ than for $Ti_I$ in the case of Ti₃AlC₂. Accordingly, the Ti-C chemical bond is stronger in Ti₂AlC than in TiC, which is also in line with the calculated Ti-C bond lengths in Table VII (2.164 Å for TiC and 2.117 Å for Ti₂AlC). When comparing the BCOOP curves of Ti₂AlC to those of Ti₃AlC₂ and TiC, it is clear that the Ti-C BCOOP curve of Ti₂AlC is more intense; this suggests that the Ti-C bond is slightly stronger in Ti₂AlC than in Ti₃AlC₂ and TiC. For Ti₂AlC, the BCOOP calculations show that the Ti $3d$ – C $2p$ hybridization and the strong covalent bonding are the origin of a low-energy carbide peak observed in the Ti $L_{2,3}$ XES spectra [27]. Finally, the Ti-Al BCOOP peak of Ti₂AlC is slightly weaker and closer to $E_F$ than in Ti₃AlC₂. This is an indication that the Ti-Al chemical bond in Ti₂AlC is somewhat weaker than in Ti₃AlC₂ as verified experimentally by the fact that the spectral weight of the peaks in the Ti $L_{2,3}$ XES spectrum is slightly shifted toward $E_F$ in Ref. [27].

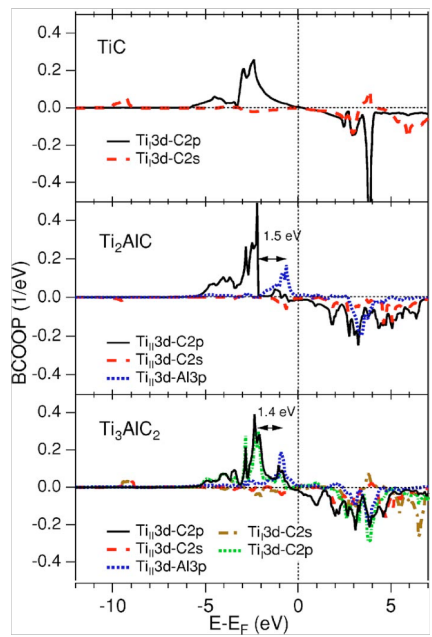

An example of a similar BCOOP analysis for a 413 phase can be found in Fig. 18, [65]. In order to explore the chemical bonding of Ti$_4$SiC$_3$, the calculated BCOOP was compared to those of Ti$_3$SiC$_2$ and TiC [65]. By inspecting the BCOOP curves and the distances of the main peaks from the Fermi level, it is shown that the Ti $3d$–C $2p$ bonds are much stronger than the Ti $3d$–Si $spd$ bonds in both Ti$_4$SiC$_3$ and Ti$_3$SiC$_2$. The Ti atoms lose some bond strength to the nearest-neighbor Si atoms, which to some degree is compensated by stronger Ti-C bonds. Furthermore, comparing the BCOOP curves of Ti$_4$SiC$_3$ to those of Ti$_3$SiC$_2$, the Ti-C BCOOP of Ti$_4$SiC$_3$ appears less intense, which indicates that the Ti-C bond is somewhat weaker in Ti$_4$SiC$_3$ than in Ti$_3$SiC$_2$. It should be noticed that the Ti$_{II}$-C$_{II}$ bonds are also shorter (2.097 Å for Ti$_3$SiC$_2$ and 2.093 Å for Ti$_4$SiC$_3$) than the Ti-C bonds in TiC (2.164 Å) as shown in Table VII. This implies that the bonds in the Ti-C slabs of the MAX phase are stronger than in TiC and are due to the weaker Ti-Si bonds which transfer charge to the Ti-C bonds.

Another detailed BCOOP analysis was made by comparing TiN, TiC, Ti$_2$AlC and Ti$_2$AlN [24]. The integrated bonding area below $E_F$ in Fig. 19 is estimated to be ~50% larger for TiC than for TiN. However, the distance of the main peak from $E_F$ is approximately two times larger in TiN compared to TiC, and this makes the covalent Ti$_{II}$ $3d$-N $2p$ bonding in TiN stronger than the Ti$_{II}$ $3d$-C $2p$ bonding in TiC. This finding is consistent to the shorter Ti$_{II}$-N bond length, computed theoretically, in Table VII (Ti-N: 2.129 Å and Ti-C: 2.164 Å). The $3d$ states in the BCOOP curves in Ti$_2$AlN are generally located further away from $E_F$ than in Ti$_2$AlC which indicates that the Ti$_{II}$-N bond is stronger in Ti$_2$AlN than the Ti$_{II}$-C bond in Ti$_2$AlC. As the Ti atoms bond stronger to N and C in one direction than to Al in the other direction, the Ti$_{II}$-N and Ti$_{II}$-C bonds are even stronger in Ti$_2$AlN and Ti$_2$AlC than the Ti$_I$-N and Ti$_I$-C bonds in TiN and TiC. This is further corroborated by the shorter bond lengths reported in Ref. [24]. The Ti$_{II}$-Al BCOOP peak at −1.1 eV in Ti$_2$AlN has a 15% larger integrated intensity than the corresponding Ti$_{II}$-Al peak at −0.64 eV in Ti$_2$AlC. This suggests that the Ti$_{II}$-Al chemical bond in Ti$_2$AlN is stronger than in Ti$_2$AlC. This is also verified experimentally by the fact that the spectral weight of the Al $L_{2,3}$ XES spectrum is stronger and slightly shifted away from $E_F$ in Ti$_2$AlN in comparison to Ti$_2$AlC.

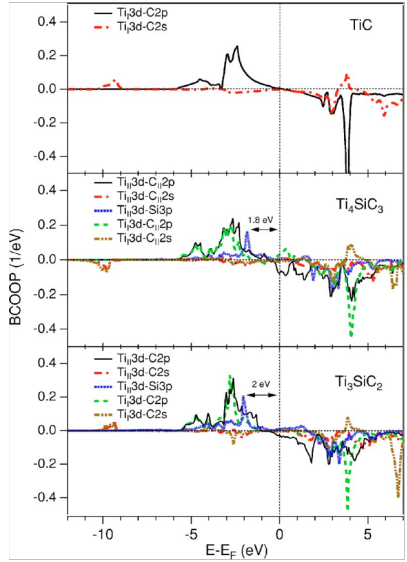

**Figure 18:** Calculated BCOOP curves for TiC, Ti$_4$SiC$_3$, and Ti$_3$SiC$_2$ *[65]*. Note that the Ti *3d* and C *2s* overlap approximately 10 eV below E$_F$ is antibonding in Ti$_4$SiC$_3$ and bonding for TiC and Ti$_3$SiC$_2$.

Figure 20 shows the chemical bonding in V$_2$GeC compared to VC investigated by calculating BCOOPs [26]. The orbital overlaps of the ternary V$_2$GeC are significantly more complex than for the binary VC system (Fig. 20). However, for both systems the main V 3*d*-C 2*p* overlap is found at −4 eV with additional peaks near −6 and −3 eV for V$_2$GeC. The V 3*d*-C 2*s* overlap has a much lower intensity than that of V 3*d*-C 2*p* with noncovalent interactions at −11 and −3 eV. The V 3*d*-Ge 4*p* overlap has a large peak at approximately −3 eV with additional smaller peaks at −3.5, −4.5, and −6 eV in V$_2$GeC. It is also noted that the V 3*d*-Ge 4*p* overlap has filled bonding orbitals up to E$_F$ while for the V 3*d*-C 2*p* overlap, antibonding orbitals start to be filled. Additionally, the integrated intensity of the V 3*d*-Ge 4*p* BCOOP curve at the −2.9 eV peak is ~16% larger than the corresponding Ti-Al peak at

−0.64 eV in Ti$_2$AlC [27]. This illustrates that the V-Ge bonding in V$_2$GeC is generally stronger than the Ti-Al bonding in Ti$_2$AlC, as highlighted by the shorter bond lengths in Table VII. Note that the V-C bond length in V$_2$GeC is also shorter than in the monocarbide VC (Table III). This interesting finding has been observed for other MAX phases [27] [24] and should play a key role in determining the mechanical properties of this class of materials (Section 10). As reported in Ref. [26], BCOOP analysis is an important tool to disentangle information about the bonding-type obtained from the study of V $L_{2,3}$ XES, C $K$ XES and Ge $M_1$ XES spectra of V$_2$GeC and VC.

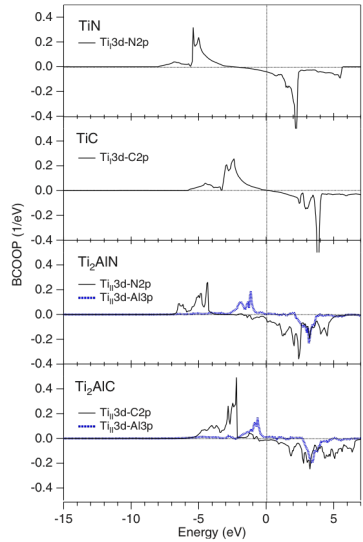

**Figure 19:** Calculated BCOOP curves for TiN, TiC, Ti$_2$AlN, and Ti$_2$AlC. From Magnuson *et al.* (2007) [24].

Generally, these special bonding characteristic trends highlighted above for several types of MAX phases are certainly influencing the physical properties of this class of layered materials, especially the elastic properties. As shown in this section, the Ti-C bond is generally stronger in MAX phases than in the parent TiC system, and the same behavior is reported for the Ti-N and the V-C bonds.

Thus, not surprisingly, this atypical bonding behavior translates into both electronic and elastic anisotropy, which is also an intrinsic property of these layered crystal structures, such as high specific stiffness values, and low Poisson ratios. When directional and localized regions of enhanced bond strengths are introduced inside a crystal, stronger directional bonds are then concentrated inside the unit cell volume, which provides low-density and stiff materials. In the same way, the presence of strong directional bonds might be responsible for the rather low Poisson values computed for MAX phases compared to their binary structures. A detailed analysis of elastic properties is given in Section 10.

## 8. Transport properties - resistivity and thermopower

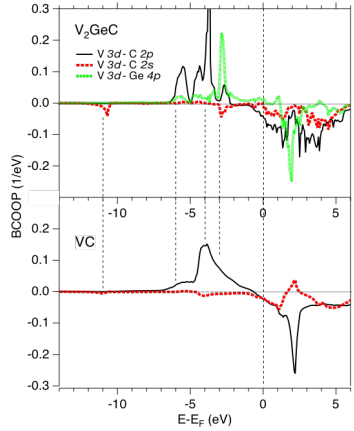

**Figure 20:** Calculated BCOOP curves for $V_2GeC$ top and VC bottom from Ref. [26].

Generally, the physical properties of MAX-phases change with constituent elements and crystal structure. For example, carbides are normally lighter and stiffer than nitrides, and the Al-containing MAX-phases are lighter and less stiff than other phases. In some cases, nitrides are better conductors than carbides.

Table VIII compares the density, resistivity and Young's modulus of the most common MAX-phases. For comparison, the density of polycrystalline Ti is 4.5 g/cm$^3$, V is 6.1 g/cm$^3$ and, Al is 2.7 g/cm$^3$. At room temperature, the resistivity of polycrystalline Ti is 0.39 μΩm, V is 197 μΩm and, Al is 26.5 μΩm.

To gain understanding about the resistivity of MAX phases, we first compare the calculated total density of states (TDOS) of the parent binary systems in Fig. 21 (top panel). For TiC, the $E_F$ is located at the bottom of a valley that indicates low conductivity i.e., high resistivity. On the contrary, for TiN and VC, the bottom of the valley is shifted toward lower energy and the TDOS at $E_F$ is much higher, indicating lower resistivity and thus higher conductivity. The resistivity is known to largely depend on the M *3d* states, while the M *4s*-states are largely suppressed at $E_F$.

**Table IV:** Macroscopic properties of selected MAX-phases in comparison to binary compounds. The computed DOS values at $E_F$ are referring the curves of Figure 21.

| Compound | Density [g/cm3] | Resistivity [μWm] | TDOS@$E_F$ [states/eV] | Young's modulus E-mod. [GPa] |
|----------|-----------------|-------------------|------------------------|------------------------------|
| TiN | 5.4 | 0.13 | 0.84 | 449 |
| TiC | 4.9 | 2.5 | 0.16 | 350-400 |
| VC | 5.8 | 0.93 | 1.14 | 255 |
| Ti$_2$AlN | 4.3 | 0.39 | 4.43 | 270 |
| Ti$_2$AlC | 4.1 | 0.44 | 2.81 | 260 (th. 305) |
| V$_2$GeC | 6.5 | 0.21 | 5.84 | 189 (th. 334) |
| Ti$_3$AlC$_2$ | 4.5 | 0.5 | 3.45 | 260 |
| Ti$_3$SiC$_2$ | 4.5 | 0.25 | 5.00 | 320 |
| Ti$_3$GeC$_2$ | 5.5 | 0.5 | 4.43 | 320 |

For the MAX phases, the situation is more complicated than in the binary parent compounds. In general, the TDOS@$E_F$ is dominated by *d-d* orbitals of the M atoms. All M$_2$AX systems (middle panel) exhibit large DOS at $E_F$ that signifies low resistivity. This is also true for the M$_3$AX$_2$ phases in the bottom panel. However, for real systems, it is not possible to determine the resistivity/conductivity only from comparison of the DOS at $E_F$ as the conductivity has a more complex behavior and therefore requires a deeper investigation taking into account, for example, mobility, defects and electron-phonon coupling. However, going beyond the general assumption of isotropic scattering (i.e., the DOS is inversely proportional to the carrier mobility) is not an easy task. This may require growing large crystals and measure the resistivity along different crystal orientations. MAX-phases are considered as *compensated conductors*, where both the number of electron and holes contribute to the conductivity in equal numbers [14].

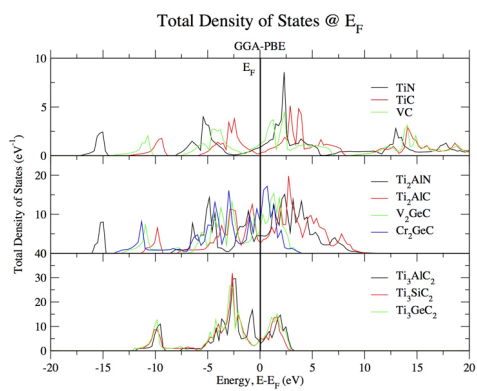

A system in which there is a particularly large discrepancy in the DOS at $E_F$ between experiment and theory is $Cr_2GeC$. This multifunctional metallic and ceramic compound exhibits a number of peculiar properties and is relatively little studied. From specific heat measurements, it has been deduced that there are 21-22 states [85] [86] at $E_F$, while DFT calculations [17] indicate that there are only 7.7 states at $E_F$. The DOS at $E_F$ was further investigated by X-ray spectroscopy at different incidence angles [25]. The agreement between theory and experiment was rather poor since the Ge *4s* states exhibit significant intensity that is not reproduced in ground-state DFT calculations at 0 K. However, this disagreement cannot be accounted for only by the effect of the rather large electron-phonon coupling, and it was found that the redistribution of intensity from the shallow *3d* core levels to the *4s* valence band provides large DOS at $E_F$. A similar disagreement between experiment and DFT results is known for Ge in $Ti_3GeC_2$ [23] and $V_2GeC$ [26] as well as for Ga in GaN [72].

**Figure 21:** Density of states at the Fermi level ($E_F$) of selected MAX phases and parent binary compounds.

The Seebeck coefficient, or thermoelectric power (μV/K), measures the magnitude of an induced thermoelectric voltage in response to a temperature difference across a material. A particular interesting property and a unique phenomenon of $Ti_3SiC_2$ is that its thermopower is zero over a wide range of temperatures (300-900 K) [87] and would potentially make $Ti_3SiC_2$ a perfect reference material in temperature measurements. This phenomenon was explained by a predicted cancellation between the partial thermopowers (Seebeck coefficients) of an occupied band along the *c*-axis and unoccupied bands in the basal *ab*-plane near $E_F$ [88]. $Ti_3SiC_2$ is also recognized as a *compensated conductor* with equal number of electrons and holes at the $E_F$. Furthermore, the highly anisotropic shear modulus in neutron diffraction exceeds theoretical predictions by a factor of three in $Ti_3SiC_2$.

In 2012, it was demonstrated by Magnuson *et al.* [16], that the in-plane Seebeck coefficient in epitaxially grown thin films of $Ti_3SiC_2$ has a positive value ranging from 4-6 μV$K^{-1}$. These results gave direct proof of an anisotropic Seebeck coefficient in single-crystal $Ti_3SiC_2$. In contrast, polycrystalline bulk samples have a negligible Seebeck coefficient. Figure 22 compares measured and calculated Seebeck coefficients for $Ti_3SiC_2$ that are in agreement with the prediction by Chaput *et al.* [88] concluding that the Seebeck coefficient is positive in the basal ab-plane but negative along the c-axis. This is also in line with $Ti_3SiC_2$ being a compensated conductor [89].

For single crystal Ti$_3$SiC$_2$, the Seebeck measurements in the basal *ab*-plane increases from 4.4 μV/K at room temperature to 6.3 μV/K at 700°C [16] and provide direct evidence of anisotropy in the Seebeck coefficient. However, S measurements along the *c*-axis (S$_{zz}$) were not possible as thin films in this orientation do not exist, but bulk polycrystalline samples exhibit S=0. Generally, calculated S$_{xx}$, and S$_{yy}$ overestimate S by ~25% at room temperature and >50% at 600 °C. The anisotropy is further overestimated in rigid band structure calculations and phonon effects must be included in the modeling.

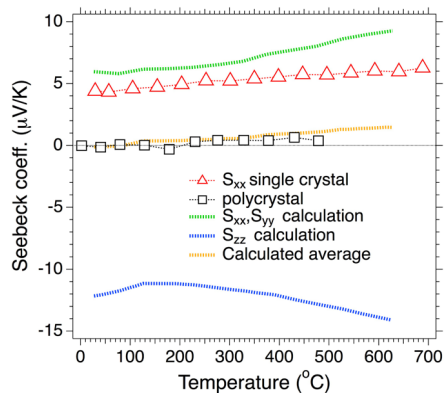

**Figure 22:** Experimental and calculated Seebeck coefficients for Ti$_3$SiC$_2$ [14].

The resistivity of MAX phases has also been evaluated using ellipsometry measurements of the dielectric function, in which it was found that free carriers contribute to the dielectric response for photon energies lower than 1.0 eV [90].

## 9. Phonons and optical properties of MAX phases: Raman and Infrared Spectroscopies

The electronic structures of MAX phases are known to be affected by phonons, i.e., particular vibrational modes and atomic moments of the different elements in the compounds. The vibrational behavior of MAX phases has primarily been studied with Raman spectroscopy [91, 92]; less is known about the changes in the electronic structure.

Figure 23 shows calculated phonon frequency spectra in the *ab*-basal plane (*x,y*) and along the *c*-axis (*z*) for the Ti$_3$SiC$_2$ phase. Phonon densities of states (PhDOS) were calculated by employing the supercell approach in the framework of DFT (GGA-PBE) and Density-Functional Perturbation Theory [93]. Specifically, a 2×2×1 supercell system was used to compute real-space force constants within the q-*ESPRESSO* software package [94]. Phonon frequencies were then obtained from the force constants using the *QHA* code [95]. As observed, there are two different overlapping regimes of phonon vibrational modes for Si with an out-of-plane high-frequency peak three to four times higher in frequency (10-12 THz) than for the peak maximum for the in-plane vibrational mode (3.3 THz). From neutron diffraction studies of MAX phases, the *A*-atoms (Si in Ti$_3$SiC$_2$) are known to act as 'rattlers' following a rather complicated ellipsoidal phonon trajectory [15]. The observed out-of-plane shift toward higher frequency for both Ti and Si indicates the presence of a structural anisotropy within the Ti$_3$SiC$_2$ unit cell. These high-frequency peaks for the z-component of Ti and Si are compensated by the lowering of the same vertical vibrational component for the C atom. Due to the overlap in the out-of-plane frequency range between Ti and Si, one can envisage a strong directional Ti-Si bonding regime along the *c*-axis. Therefore, the complex rattling behavior found experimentally for the Si atom can be verified by the large frequency difference between the out-of-plane and in-plane vibration.

When investigating the calculated PhDOS within the 312-series $Ti_2SiC_2$, $Ti_2AlC_2$ and $Ti_2GeC_2$, it is observed that the lowest vibration energies are always those of the *A*-elements, pointing to the aforementioned rattling behavior. This shaking nature of the A atoms was also corroborated by their higher amplitudes of vibration with respect to the *M* or *X* species. The lower in-plane vibrational energies for the *A*-layers also explain why the A-atoms tend to vibrate more along the basal plane than along the [0001] direction. Studies of the atomic displacement parameters further confirm that vibrations of the A-elements are more prominent along the basal plane of the crystal than normal to the plane [15]. In agreement with Raman studies, vibration of the C atoms occurs at higher energies, while the vibrations of the heavier Ti atoms are between those of A and C atomic species. A complex (ellipsoidal) correlated motion between Ti and the A elements has also been proposed theoretically [96].

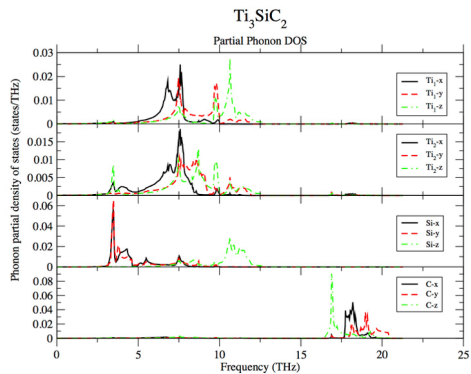

**Figure 23:** Calculated phonon DOS of $Ti_3SiC_2$.

A number of different Raman-active vibrational modes have been experimentally and theoretically determined for a variety of MAX phases [92] [97] [98] [99] [100] [101] [102]. In particular, two types of Raman active modes were identified in all MAX phases. The first branch, between 50 and 300 $cm^{-1}$, refers to the low energy modes attributed to vibrations of the A and M atoms. Two higher energy modes (550 and 650 $cm^{-1}$) due to the vibrations of the X atoms were only observed in the 312 and 413 phases. In general, *first-principles* calculations agree well with these experimental results [97] [102] [98] [99] [15]. For the 211 phases, four principal Raman frequencies ($\omega_1$ to $\omega_4$) were identified. From the Raman spectrum of the $V_2AlC$ single crystal, Spanier *et al.* [97] [91] also showed that these modes involve M and A atoms, whereas C plays only a minor role. Modes $\omega_1$, $\omega_2$ and $\omega_3$ correspond to vibrations in the basal planes, whereas the highest energy mode $\omega_4$ is due to atomic vibration along the [0001] direction. Note that the characteristic mode $\omega_3$ has basically the same frequency ($\approx$300 $cm^{-1}$) in both $Ti_2AlC$ and $Ti_3AlC_2$, i.e., not a strong dependence on crystal structure. Raman spectra for the 312 phases (*e.g.*, $Ti_3SiC_2$, $Ti_3AlC_2$ and $Ti_3GeC_2$) show four characteristic modes at 190-200, 279-297, 625-631, and 664-678 $cm^{-1}$ [102]. Here, the interaction between X atoms in the MX layers give rise to the highest frequency modes in both 312 and 413 systems, while the lower frequencies are due to the A-group elements.

Using *ab initio* calculations, it is possible to predict trends of measured modes *versus* the reduced mass. However, theoretical energy values are often computed to be lower than the experimental ones by 10-25%. First-order Raman spectra for the 413 phases were both measured and predicted by DFT for $Ti_4AlN_3$ [97], $Nb_4AlC_3$ [15] and $Ta_4AlC_3$ [101]. In these phases, ten Raman active modes ($3A_{1g}+3E_{1g}+4E_{2g}$) were highlighted with good agreement between theory and measurements, and then used to infer the M-X bond stiffness. The achieved sequence in bond strength (Ta-C > Nb-C > Ti-N) was found to be in accordance with the experimental trend in melting points (3983° for TaC, 3600° for NbC and 2949° for TiN [103]).

Measurements using infrared spectroscopy have been sparse and so far, only one focused study has been reported for $Ti_3GeC_2$ [104]. In this work, four of the five expected infrared modes were observed for this material. Understanding of the optical properties of MAX phases is of importance for comprehending the fine electronic structure of this class of materials. In fact, for MAX phases that are metallic-like conductors, the optical properties are functions of the delocalized electron polarization and the inter-band transitions. Few experimental studies on the optical response of MAX phases have been reported and compared with *ab initio* calculations for $Ti_3SiC_2$ and $Ti_4AlN_3$ [105], $Ti_2AlC$ and $Ti_2AlN$ [106], $Ti_3AlC_2$ [107], $Ti_2AlN$, $Ti_2AlC$, $Nb_2AlC$, $TiNbAlC$, and $Ti_3GeC_2$ [90]. A number of theoretical papers dealing with the optical response of MAX phases have been published [108] [109] [110] [111], providing an interesting amount of optical data to be compared with further measurements. Lastly, following the theoretical work of Li *et al.* in 2008 [105] on $Ti_4AlN_3$, attention has been given to the study of reflectivity spectra of MAX phases. In fact, Li et al. concluded that $Ti_4AlN_3$ could reduce solar heating and enhance the infrared emittance, thus moderating the equilibrium surface temperature of a MAX-phase-coated spacecraft.

## 10. Elastic properties of MAX-phases

MAX phases represent a class of stiff materials (*e.g.*, $Ti_3SiC_2$ is comparable to $Si_3N_4$ and $Hf_3N_4$ [112] in stiffness) with low-density values ($\approx$4-5 g/cm$^3$). It is this special combination that makes their specific stiffness high. Considering that they are also readily machinable with normal high-speed tool steels and are damage tolerant, this group of materials possesses a unique combination of mechanical properties [113]. Worth noting is that machining does not occur by plastic deformation of the material, as in the case of metals, but rather through the breaking of small microscopic flakes. These unusual mechanical properties can be traced to their intrinsic layered nature (*i.e.*, they are plastically anisotropic) and crystal symmetry. MAX phases have a hexagonal crystal symmetry that implies the existence of five independent elastic constants ($C_{11}$, $C_{12}$, $C_{13}$, $C_{33}$, and $C_{44}$). For sake of coherence in their trends, the elastic constants were computed in this review (Table V) for a number of representative phases employing the same computational scheme (Medea-VASP [114]). Discrepancies in computed elastic properties are often observed and usually assigned to the approximations made in the various computational schemes used by the different authors [115, 115]. For of more complete appraisal of the theoretical elastic properties in more than 240 MAX phases, the reader is referred to the works of Cover *et al.* (2009) [28] and Barsoum (2013) [1].

Unfortunately, due to the lack of large single-crystal bulk samples, a direct measurement of the elastic constants in MAX phases is basically absent. This explains why *ab initio* estimate are of special importance. Nonetheless, when grains are small and the sample deforms linearly, as in the case of the $Ti_2SC$ phase, stress-strain curves can be obtained via an in-situ neutron beam [116]. In this case, an indirect way of comparing theoretical and experimental $C_{i,j}$ values is achieved through the elasto-plastic self-consistent approach [117] [118] [119] [120], which uses the DFT-computed elastic constants as input. The model treats grains as they were inclusions in an infinite and homogeneous matrix, whose elastic properties are similar to the average polycrystalline moduli of all grains in the sample. Thus, since the model averages over a large number of grains, the large sampling volume of the neutron diffraction measurements can be captured. Applying this methodology to $Ti_2SC$, it is found that *ab initio* elastic constants are valid, and have an accuracy of $\approx 10\%$. Considering that theoretical values were obtained at 0 K and measurements were instead carried out at ambient temperatures, the achieved results suggest that DFT calculations for $Ti_2SC$ are reasonably accurate for most practical purposes. However, for the rest of MAX phases, until direct measurements of the elastic constants are available on large single crystals, *first-principles* calculations remain the main tool to study coherent trends in the elastic properties. Temperature effects further complicate these investigations. As observed in most solids, an increasing temperature in MAX phases leads to a reduction in the elastic moduli [78].

**Table V:** Theoretical elastic constants and averaged polycrystalline properties for various MAX-phases. The calculations were carried out for relaxed PBE geometries using the MedeA-VASP software package [84]. The Elastic constant ($C_{ij}$) and the averaged moduli are given in GPa, sound velocities in km/s, and the Debye Temperature ($\Theta_D$) in K.

| System | $C_{11}$ | $C_{12}$ | $C_{13}$ | $C_{33}$ | $C_{44}$ | $B_H$ | $G_H$ | $E_H$ | $V_s$ | $V_p$ | $\Theta_D$ | $\nu$ | $G_H/B_H$ |
|---|---|---|---|---|---|---|---|---|---|---|---|---|---|
| TiN | 615.7 | 149.7 | - | - | 167.8 | 305.0 | 191.4 | 474.9 | 5.84 | 10.17 | 927.2 | 0.25 | 0.63 |
| TiC | 524.5 | 121.9 | - | - | 171.2 | 256.1 | 182.7 | 442.8 | 6.10 | 10.10 | 927.9 | 0.21 | 0.71 |
| VC | 642.8 | 126.8 | - | - | 181.9 | 298.8 | 209.3 | 509.0 | 5.98 | 9.94 | 949.2 | 0.22 | 0.70 |
| $Ti_3AlC_2$ | 360.6 | 75.9 | 74.2 | 305.0 | 147.3 | 163.4 | 140.5 | 327.6 | 5.76 | 9.11 | 808.3 | 0.17 | 0.86 |
| $Ti_3SiC_2$ | 378.6 | 84.2 | 100.4 | 361.0 | 172.0 | 187.6 | 152.9 | 360.7 | 5.83 | 9.33 | 834.3 | 0.18 | 0.82 |
| $Ti_3GeC_2$ | 358.8 | 80.0 | 89.5 | 337.9 | 154.8 | 174.8 | 142.5 | 336.1 | 5.14 | 8.23 | 729.9 | 0.18 | 0.82 |
| $Ti_2AlC$ | 304.8 | 53.4 | 51.6 | 260.8 | 128.5 | 131.2 | 123.9 | 282.7 | 5.56 | 8.60 | 754.4 | 0.14 | 0.94 |
| $Ti_2AlN$ | 316.8 | 60.6 | 77.9 | 267.3 | 138.7 | 148.0 | 125.8 | 294.1 | 5.40 | 8.55 | 749.1 | 0.17 | 0.85 |
| $V_2GeC$ | 291.0 | 98.9 | 130.7 | 296.4 | 159.2 | 177.3 | 112.5 | 278.5 | 4.15 | 7.09 | 601.0 | 0.24 | 0.64 |
| $Ti_4SiC_3$ | 401.6 | 84.6 | 103.3 | 384.3 | 176.4 | 196.7 | 161.4 | 380.3 | 5.94 | 9.49 | 860.4 | 0.18 | 0.82 |

Not surprisingly, all computed $C_{ij}$ values in Table satisfy the Born and Huang's stability criteria [121] for hexagonal (MAX-phases) and cubic (MX binary systems) crystal symmetries. From the estimated Hill's [122] averaged bulk ($B_H$), shear ($G_H$) and Young's ($E_H$) moduli, also shown in Table V, it is evident that the investigated MAX phases are all softer than their parent binary systems. However, as a general tendency, Si-containing phases ($Ti_3SiC_2$ and $Ti_4SiC_3$) are significantly harder than $Ti_2AlC$ and $V_2GeC$ phases. Furthermore, for all the MAX phases considered in Table V, both the shear modulus and the $C_{44}$ elastic constant value is always smaller than the bulk modulus. This is consistent with the fact that MAX phases have a better damage tolerance than most other ceramic systems.

The resistance of a material to an applied hydrostatic pressure can be measured by the magnitude of the bulk modulus. Many factors, such as unit cell volume, number of valence electrons, defects, and puckering of the basal planes can affect this value in MAX phases. Intuitively, for isostructural compounds, the smaller the unit cell volume, the stronger the bonds, and therefore, the higher the bulk modulus. When correlating the theoretical bulk moduli with the electronic configuration of the *A*-element for various $Ti_2AC$ phases [74], it is found that *B* increases when the *A*-group element increases from 13 to 15. In the same manner, the computed bulk moduli for all known 211 phases [28] show a good correlation with the unit cell volume, although such correlations are less evident in experiments. However, it is still unknown why for a given unit cell volume, the larger the *M*-atoms, the harder is the resulting solid. Not surprisingly, various theoretical results [123] [78] [124] have shown a direct correspondence between the bulk moduli of the MAX-phases and those from the parent MX binary compounds. The latter correlation has been confirmed experimentally. Another important factor that influences the experimental bulk modulus is the vacancy concentration that tends to reduce the value of $B$, thus making a comparison with the theoretical $B$ value obtained from a pure vacancy-free crystal problematic. Corrugation of the basal planes can also affect the value of the bulk modulus. Using EELS together with *ab initio* calculations, it has been shown [69] that the basal planes in TiNbAlC solid solution are corrugated. This kind of puckling effect relates to softening of the crystal along the *c*-axis which can account for a significant decrease in the bulk modulus. In general, making reference to the measured $B$ values to validate *first-principles* calculations, or the other way around, is a very difficult task, as many pitfalls are known to exist in both directions. Discrepancies stem from two main sources, (*i*) different types of approximations made in theoretical calculations (e.g., GGA *vs.* LDA [125]), and (*ii*) the fact that real samples are not defect-free and measurements are carried out at ambient temperatures. For the latter effect, it is well known that increasing the temperature reduces the elastic moduli of the material. However, from the large experience accumulated in comparing theoretical and experimental trends, a widespread strategy to engineer the bulk modulus in MAX phases has been delineated. For example, in order to minimize $B$, one has to choose a MAX phase whose parent MX phase has the highest $B$ value, and at the same time maximize the number of electrons.

The Poisson ratio (ν) shown in Table  has been estimated using theoretical sound velocities ($V_p$ and $V_s$) [126]. ν is the ratio of the transverse to longitudinal strain for an elastically-loaded material. Hence, it quantifies the stability of the crystal against shearing. Poisson's ratio can formally take values between −1 and 0.5, which corresponds, respectively, to the lower bound where the material does not change its shape and to the upper bound when the volume remains unchanged. For systems with predominantly central interatomic interactions (*i.e.*, ionic crystals), the value of ν is usually close to 0.25, which is the reference value for a perfectly isotropic elastic material. Thus, ν decreases as non-central effects become more important. For most of the MAX phases, computed Poisson ratios are found to be approximately 0.20 or slightly lower, which means that they are affected by non-central (covalent-like) forces. The lowest ν value (0.14) was calculated for $Ti_2AlC$, pointing to an enhanced directional bonding character for this phase. In contrast, $V_2GeC$, and $Cr_2AlC$ [127] have ν = 0.24 suggesting a more isotropic elastic behavior. The highest Poisson ratio (0.29) was computed for $Cr_2GeC$ [124], a result which is very much close to that of Ti, steels and most other metals (*i.e.*, ν = 0.3).

Using the computed $G_H/B_H$ ratios of Table V, one can further estimate the brittle and ductile behavior of polycrystalline MAX phases by considering $B_H$ as the resistance to fracture and $G_H$ as the resistance to plastic deformation. Therefore, a high (low) $G_H/B_H$ ratio becomes associated with brittleness (ductility). The critical ratio that separates ductile and brittle was estimated at about 0.57 [128]. As a general tendency, we see that MAX phases are computed to be more brittle than their reference binary compounds.

One of the most important parameters that determines the thermal characteristics of materials is the Debye temperature ($\Theta_D$). Since this temperature corresponds to the temperature of a crystal's highest normal mode of vibration, it can be used to correlate the elastic properties with phonons, thermal expansion, thermal conductivity, specific heat and enthalpy. As a rule of thumb, a higher $\Theta_D$ implies higher thermal conductivity. Thus, knowledge of $\Theta_D$ is essential for developing and manufacturing electronic devices. In this review, we made use of the simple Debye-Grüneisen model [129] to estimate the magnitude of $\Theta_D$ for the investigated MAX phases. The Debye temperature can be defined in terms of the mean sound velocity and gives explicit information about lattice vibrations [130]. Our fi*rst-principles* calculations suggest that MAX phases have a Debye temperature lower than reference binary compounds, pointing to a less stiff lattice and therefore a potentially lower thermal conductivity. However, $\Theta_D$ values are all above 600 K suggesting that MAX phases are still hard compounds with a large wave velocity and therefore a relatively high thermal conductivity. Although the agreement between $\Theta_D$ values from different computational schemes can sometimes be considered acceptable, there exist cases where the estimated values can differ by more than 200 K (for $Ti_3SiC_2$ *cf.* the value reported in Table IV and those in references [131] [132] [131]). Therefore, these numbers should be taken in a relative way, to obtain trends within different MAX-phases, while avoiding conclusions based on their absolute values.

As is seen from the computed $E_H$ values in Table V, both $Ti_3SiC_2$ and $Ti_4SiC_3$ have the highest Young modulus (~360-380 GPa) among the series of selected MAX phases. This result corroborates previous results that pointed to a high plasticity for the $Ti_3SiC_2$-based materials [133] [134]. From the computed electronic structure, it has been shown that the weak interaction between the layers containing $Ti_6C$ octahedra and networks of Si atoms explains such a high plasticity [80]. Agreement between measured and calculated young's moduli is good. However, the theoretical values tend to be larger than the experimental numbers, probably because of the presence of defects (vacancies and impurities) in the sample that decrease the elastic properties and because DFT calculations are performed at 0 K.

Trends in computed $E$ and $G$ can be found when looking at the results in Table V as a function of the average number of valence electrons and the size of the $A$-group element. In the first case, when too many electrons have to be accommodated, a significant number of them will fill up anti-bonding orbitals generating instability in the crystal. This is probably the case for $Cr_2GeC$, where the large number of average valence electrons (5) is responsible for a sudden drop of $E$ and $G$ among various MAX phases. In fact, a rather low $G$ value has been measured (80 GPa [135]) and theoretically corroborated (80.5 GPa [28], and 96.7 GPa [136]). The same behavior has also been observed for the Young's modulus, with an experimental value of 208 GPa [78] and a theoretical value between 213 GPa [124] and 249 GPa [69]. The large instability for $Cr_2GeC$ is consistent with its high DOS@$E_F$ and large thermal expansion. The size of the $A$-group element further affects the magnitudes of both $E$ and $G$, with an almost linear decrease in these values with an increase in the $A$-radius. Systematic changes in both $E$ and $G$ as a function of the $M$-group element have been predicted theoretically [137] [55] [138] [139], even though no significant experimental correlation has been reported. The effect of $X$ seems to be weak and not very important in tuning the elastic properties of MAX phases. Therefore, in order to engineer the elastic properties of MAX phases to obtain stiff materials, the rule of thumb is to choose small atoms (especially for the $A$-element) and maintain the total number of valence electrons as low as possible. Following these suggestions, Cover *et al.* (2009) [28] predicted theoretically the highest $E$ and $G$ values for $V_2PC$, extending the search among the entire set of known 211 MAX phases.

**Table VI:** Calculated charge-transfer relative to atoms.

| System | $M_1$ | $M_2$ | A | $X_1$ | $X_2$ |
|--------|-------|-------|---|-------|-------|
| TiN | 1.4125e | | | -1.4125e | |
| TiC | 1.2554e | | | -1.2554e | |
| VC | 1.1236e | | | -1.1236e | |
| Ti$_3$AlC$_2$ | 1.1058e | 0.8064e | -0.1742e | -1.2722e | |
| Ti$_3$SiC$_2$ | 1.6110e | 1.4253e | -0.9175e | -1.7721e | |
| Ti$_3$GeC$_2$ | 1.6011e | 1.4293e | -0.8746e | -1.7926e | |
| Ti$_2$AlC | 0.6414e | | -0.1105e | -1.1723e | |
| Ti$_2$AlN | 0.8988e | | -0.8392e | -0.9585e | |
| V$_2$GeC | 1.0022e | | -0.4959e | -1.5084e | |
| Ti$_4$SiC$_3$ | 1.6323e | 1.4327e | -0.9114e | -1.7177e | -1.7504e |

The elastic anisotropy of crystals is an important parameter for engineering science since it correlates with the appearance of micro-cracks and structural defects during crystal growth and/or manufacturing processes [140]. As mentioned earlier, in a hexagonal lattice, there are five independent elastic constants, and the elastic anisotropy is described in terms of one compressional $\Delta_P = C_{33}/C_{11}$ and two shear anisotropy ratios, $\Delta_{S1} = (C_{11}+C_{33}-2C_{13})/4C_{44}$ and $\Delta_{S2} = 2C_{44}/(C_{11}-C_{12})$. Hence, the hexagonal lattice is isotropic if $C_{11}=C_{33}$, $C_{12}=C_{12}$ and $C_{11}-C_{12} = 2C_{44}$. For crystals with isotropic elastic properties $\Delta_P = 1$, $\Delta_{S1} = 1$, and $\Delta_{S2} = 1$, while values smaller or greater than unity provide a measure of the degree of elastic anisotropy. Using the set of computed $C_{ij}$ values from Table V, it is found that $\Delta_P$ is generally close to unity except for Ti$_3$AlC$_2$, Ti$_2$AlC, and Ti$_2$AlN for which the value is 0.85. For most of the investigated MAX phases, the first shear ratio ($\Delta_{S1} = 1$) is close to 0.80, while for V$_2$GeC it is computed to be exceptionally high (1.65). This elastic anisotropy in shearing is further highlighted in the value of $\Delta_{S2}$ which is close to unity for most of the studied phases, but is drastically lower, 0.51, for V$_2$GeC.

## 11. Charge-transfer in MAX-phases

DFT calculations indicate that there are trends in charge-transfer from the metal *3d* and *4s* orbitals in both MAX- and MX-phases that depends on the electronegativity of the constituent elements, as shown in Table VI. For the binary MX compounds, the charge transfer is thus larger for nitrogen than for carbon, and in general is significantly larger for the Ti-containing systems than those having V. The main charge reservoir resides on the *3d* orbitals of the M-element, while a less extensive amount originates from the *4s* orbitals.

In the case of MAX phases, the calculations indicate charge transfer to all the A-elements and are estimated to be higher in the Si- and Ge-containing phases, such as Ti$_3$SiC$_2$, Ti$_3$GeC$_2$ and Ti$_4$SiC$_3$. The large charge density variations can be taken as a sign of the ionic bonding character of the Ti-Si and Ti-Ge bonds. Moreover, the gain in electronic charge at the Si (or Ge) site may further explain the phonon rattling behavior of the A-atoms. As a general tendency, we observe that the transition metal atoms always lose electronic charge (positive charge transfer values in Table VI), while C or N and the *A*-element tend to gain charge (values in Table VI are negative). Although the set of charge transfer values clearly depend on the employed computational method [69], common agreement in the above general tendency has been achieved. For Ti$_2$AlC, an indirect experimental indication of such a charge transfer from Ti to C and Al (as listed in Table VI) is the measured Ti *2p$_{3/2,1/2}$* core-level XPS values. Magnuson *et al.* (2006) [27] reported that in Ti$_2$AlC there is a high-energy shift in the binding energies due to screening compared to pure Ti. Moreover, by following the same screening criteria, they found that the increased Al charge density is further corroborated by the measured XPS binding energies of Al in $Ti_2AlC$, which are shifted to lower energy (72.5 eV) compared to pure Al (72.8 eV). This effect is even more pronounced for C (281.9 eV) in comparison to amorphous C-C carbon (284.8 eV). A very similar trend in the chemical shift was detected for XPS-binding energies in $Ti_3AlC_2$ [23].

## 12. Magnetic properties of MAX phases

The potential for magnetism in MAX phases is an important issue and an active field of research since this would add a new property to a large number of phases that can be useful in various applications. Since magnetic ordering could also affect other properties of the MAX-phases such as thermal expansion and the bulk modulus, it is important to know the atomic magnetic exchange interactions in detail as non-magnetic calculations yield insufficient agreement with experiment.

In 2004, different magnetic configurations; ferromagnetic (FM), antiferromagnetic (AFM) and nonmagnetic (NM), in $Cr_2AlC$ were considered theoretically by Schneider *et al.* [141]. Later, Luo *et al.* [142] investigated the hypothetical Fe-containing 211 MAX-phases. However, more accurate calculations including all competing phases such as inverse perovskites [143] and quaternary phases showed that the Fe-containing 211 system was thermodynamically unstable [144]. Experimentally, a magnetic signal was later observed in $Cr_{2-x}Mn_xGaC$ phases [145], and attributed to the presence of magnetic perovskite impurities. In Mn-doped $(Cr,Mn)_2GeC$ at room temperature, the magnetic signal was more clearly observed [146]. In particular, for $Cr_2GeC$, the presence of two competing magnetic mechanisms (FM in plane and AFM along the *c*-axis), can lead to stabilization of a non-perfectly compensated AFM material if the non-collinearity of the spins is fully taken into account. Employing DFT+U full-potential calculations with different exchange-correlation functionals, Mattesini *et al.* [17] showed that $Cr_2GeC$ is a weak AFM material.

The nature of the correlation effects of localized Cr *3d* states and the competing balance between FM and AFM ordering with a very small energy difference, make modeling of magnetic coupling complicated [73] [147] [148]. In element-specific experimental data, it is therefore of special interest to assess the origin of the magnetic coupling between the Cr atoms [141] [25] [149]. Here, the $Cr_2GeC$ phase can be used as a model test case for other similar magnetic MAX phases such as, for example, $Cr_2AlC$ [141], and $V_2GeC$ [60].

Macroscopically, $Cr_2AlC$ is non-magnetic at room temperature, but the Cr atoms can exhibit low-temperature magnetic ordering (FM, AFM, NM, or paramagnetic (PM)) [25]. This is evidenced by the X-ray Magnetic Circular Dichroism (XMCD) work of Jaouen *et al.* [147] who discovered that Cr *3d* states in $Cr_2AlC$ and $Cr_2GeC$ are FM coupled at very low temperature (2.2 K) and high magnetic fields (10 T). However, XMCD measurements of MAX phases are complicated for several other reasons, not only sample phase purity, but also magnetic surface oxides that are quickly formed even on relatively fresh samples, as well as very weak magnetic signals that require long measurement times. It is worth mentioning that the observed FM coupling in $Cr_2GeC$ has not yet been clearly corroborated by theoretical studies.

Using standard DFT-GGA, it was found that the ground state of $Cr_2GeC$ at 0 K is AFM and the ferromagnetic (FM) configuration is only a metastable state [148]. In the AFM case, a significant spin-splitting of ~2 eV was predicted by Zhou *et al.* [148] at the Fermi level. However, using the same GGA functional, Ramzan *et al.* [73] predicted a NM ground state for $Cr_2GeC$, thus emphasizing the general theoretical difficulties to converge to a well-accepted magnetic ordering. By introducing the onsite Coulomb repulsion (+$U$) to the localized $3d$ electrons, more recent $GGA^{WC}$+U calculations for $Cr_2GeC$ by Mattesini *et al.* [17] and Magnuson *et al.* [25] pointed out that the magnetic moments on the Cr atoms, although in-plane FM coupled, are small and largely cancel each other along the $c$-axis. The latter vertical coupling could lead to a perfect AFM ordering or to a ferrimagnetic ground-state if a spin collinear effect is taken into account. The Heyd-Scuseria-Ernzerh (HSE06) hybrid functional formalism [150] has been employed for a variety of different spin configurations and also provides an AFM ground-state ordering [73] [60] [151]. Furthermore, by making use of expanded unit cells (i.e., the supercell approach [152]), in order to remove possible size constraints, more complex magnetic ordering has been investigated for $Cr_2AC$, with A=Al, Ge, and Ga. This alternative method provided for $Cr_2AlC$ a ground-state magnetic ordering with an in-plane AFM spin configuration [153]. Hence, besides the existing controversial theoretical interpretations about which type of AFM coupling dominates (i.e., in-plane or out-of-plane), the evident disagreement between theory (AFM) and experiment (FM) in determining the lower energy magnetic ordering in $Cr_2GeC$ calls for the need of further experimental and theoretical efforts [73] [154] [17] [115] [155]. Experimentally, an alternative and more bulk-sensitive spectroscopic technique than XMCD [50], such as resonant magnetic X-ray scattering (XRMS) [156] utilizing constructive interference at suitable Bragg scattering angles [21] would be more useful for measuring weak magnetic signals in nanolaminates.

In order to enhance the magnetic properties of MAX phases, alloying with Mn has also been investigated by Dalhqvist *et al.* in 2011 [144] who predicted theoretically the existence of a FM $(Cr_{0.5}Mn_{0.5})_2AlC$ phase that stimulated bulk synthesis of various Mn alloyed $Cr_2AlC$ systems. A first attempt was performed by Mockute *et al.* [157], who also worked on a magnetron-sputtering-synthesized $(Cr_{0.84}Mn_{0.16})_2AlC$ film sample and found a soft FM behavior. Later, more attempts to stabilize $Cr_2AlC$, $Cr_2GeC$, and $Cr_2GaC$ with significant additions of Mn were also carried out [146] [158]. However, the magnetic measurements showed no FM coupling or any type of magnetic transitions up to 300 K. In the case of $(Cr_{0.75}Mn_{0.25})_2GeC$, the calculated magnetic ground state was found to be AFM, but with the FM coupling very close in energy. Experimental results showed a magnetic signal up to room temperature with a very small remanence detected up to 200 K, suggesting a competition between different pure magnetic states (AFM *vs.* FM).

Many material compositions were studied so far and, in general, these potentially magnetic MAX phases have exhibited weak ferromagnetism at low temperatures with some exceptions where the transition temperatures were found above room temperature [159]. Thus, the nature of the magnetic behavior of magnetic MAX phases seems to be complex and difficult to explain theoretically. At the moment, different explanations have been proposed, ranging from itinerant electron magnetism, Pauli paramagnetism, spin density wave states to non-collinear spin ordering. The present large discrepancy between the measured and computed magnetic moments, where the measured moments are often significantly lower than calculated ones, may not only be due to different sample quality, but also to diverse types of synthesis routes. However, the divergence between theory and experimental observations is an indication of a new field of research that is rapidly developing towards the discovery of novel phases with intriguing magnetic properties.

## 13. Concluding remarks

As shown in this critical review, the physical properties of MAX phases are complex in their nature and combine the best of ceramics and metals. Information obtained about chemical bonding, the symmetry and the anisotropy in the electronic structure, can be obtained using polarization and angle-dependent X-ray absorption and emission spectroscopies. The emission spectral features are also sensitive to the local coordination of the crystal structures in- and out-of-plane. Generally, there are three different types of bonding; relatively weak M *3d* – A *3p* bonding close to the Fermi level, as well as M *3d* - C *2p* and M *3d* - C *2s* bonds, that are stronger and deeper in energy than the Fermi level. The electronic structure and chemical bonding is significantly different in the nitrides than in the isostructural carbides. In fact, there exists a relationship between the macroscopic properties and the corresponding $M_{n+1}AX_n$ bond strength in carbides and nitrides. The electronic structures of the ternary and binary nitrides were found to be significantly different than the isostructural carbides with M-X bonding states located at deeper energies. This leads to stronger bonding and higher specific stiffness values. More generally, a modification of the chemical bond strengths can also be achieved by exchanging the A-element which results in a change of the valence electron population (e.g., Al to Si, and Ge in MAX-phases) or changing the crystal structure (211 to 312, 413). This means that the chemical bond length/strength and materials properties can be tailored for specific applications.

There is a substantial anisotropy in the electronic structure of the 211 phases $V_2GeC$ and $Cr_2GeC$, as well as the 312 phase $Ti_3SiC_2$ that is correlated to transport properties such as conductivity and thermopower. In $Ti_3SiC_2$, there is experimental evidence of anisotropic Seebeck coefficients as the measured Seebeck coefficient in the basal *ab*-plane is positive and increases with temperature. The greater number of in-plane states at $E_F$ gives a positive contribution to the Seebeck coefficient in the basal *ab*-plane, but is compensated with a negative contribution along the *c*-axis so that the related electron- and hole-like bands at the Fermi level yield an average Seebeck contribution of zero. As a general tendency, the A-elements are highly influenced by phonons which have higher frequency along the *c*-axis than in the basal *ab*-plane. This gives rise to a spectral anisotropy and provides a better explanation for the temperature dependence of the Seebeck coefficient when included in rigid-band electronic structure calculations. All MAX phases are good conductors with high density of states at the Fermi level, and relatively low Seebeck coefficients. However, for $Cr_2GeC$, the exceptionally high density of states at $E_F$ is related to an intensity redistribution from the shallow Ge *3d* core level to the *4s* valence states. Thus, the greater intensity of the Ge *4s* states observed experimentally explains the large difference between the experimental observation and the calculated DOS at $E_F$. Another example of complexity in the physical properties of MAX phases is the fact that thermal expansion, elastic properties, and thermal conductivity largely resemble those of the parent MX binaries, whereas the electronic structure and transport properties are more similar to those of the parent transition metals.

Outstanding questions for the future are anticipated applications of MAX phases in the form of thin films and bulk materials. Tuning conductivity properties from metal toward semiconducting behavior may be accomplished by doping. MAX phases are particularly suited for applications in engine parts, heating elements in melting furnaces, low-friction electrical contacts to avoid arcs, and as coating materials in rockets. In the demanding environment as *electrical contacts* (pantographs sliding along catenaries) for high-speed trains, initial tests of $M_{n+1}AX_n$-phases of $Ti_3SiC_2$ and $Ti_3AlC_2$ were found to perform better than graphite-based materials. More information is necessary for unexplored and exotic properties such as magnetism, optical properties and superconductivity of MAX-phases. In particular, anisotropic measurements associated with electrical, magnetic, and thermopower properties require more experimental and theoretical efforts on MAX-phase electronic structure and chemical bonding. The effects of impurities and vacancies on the physical properties of MAX phases also represent an important research direction that has not yet been fully scrutinized.

A strategic key point for future development of MAX phases is the synergistic relationship between theoretical and experimental methods, which is nowadays a trustworthy recipe for discovering and designing new compounds able to meet challenging technological needs. Considering the vast number of already theoretical-predicted MAX phases, it is reasonable to expect that many more unexplored and exotic phases are still left to be revealed. However, while continuing with this key search process, central and highly debated issues on the existence of missing MAX-phases, such as $Ti_3AlN_2$ and $Ti_2SiC$, should be undertaken. Furthermore, along with the pursuit for novel thermodynamically-stable stoichiometries, more efforts are required in order to shed light on superconductivity in MAX phases, nanotube-structured systems, and novel magnetic phases (e.g., magnetic MXenes, and ordered $(V,Mn)_3GaC_2$ systems [159] [160] [161]). Tuning conductivity properties in MAX phases by doping from metal to semiconductor also deserves a more specific theoretical consideration and, above all, many more experimental attempts. Additional momentum-dependent measurements and theoretical reference data are also needed to obtain deeper insights on the anisotropy in the electronic properties of these phases. Lastly, of special interest is the current development of modern synthesis and materials processing techniques, which can render MAX phases corrosion and oxidation resistant, and with relatively low friction. They can also be made resistant to thermal shock, while retaining their strength in temperatures above 1300 degrees C in air. Due to these unique characteristics, the nanolaminated and nanocomposite modifications of $Ti_3SiC_2$ have already found interesting commercial applications in several worldwide companies, including, Sandvik/Kanthal, Seco Tools, IonBond, ABB, Impact Coatings, and Volvo AB.

# Acknowledgements

We gratefully acknowledge all valuable discussions in the MAX phase community. Martin Magnuson acknowledges financial support from *the Swedish Foundation for Strategic Research (SSF)* (no. RMA11-0029) *through the synergy grant FUNCASE* and the Carl Trygger Foundation. Maurizio Mattesini acknowledges financial support by the *Spanish Ministry of Economy and Competitiveness* (*CGL2013-41860-P*), and by the *BBVA Foundation* (PR14 CMA10) under the "*I convocatoria de Ayudas Fundación BBVA a Investigadores, Innovadores y Creadores Culturales*".

## References

[1]   M. W. Barsoum, MAX Phases: Properties of Machinable Ternary Carbides and Nitrides, Wiley, 2013.

[2]   M. W. Barsoum, "Dislocations, kink bands, and room-temperature plasticity of Ti3SiC2," *Metall. Mater. Trans. A*, vol. 30, p. 1727, 1999.

[3]   W. Jeitschko, H. Nowotny and F. Benesovsky, "Ein Beitrag zum Dreistoff Molybdän-Aluminium-Kohlenstoff," *Mh. Chem.*, vol. 94, pp. 247-251, 1963.

[4]   W. Jeitschko, H. Nowotny and F. Benesovsky, "Kohlenstoffhaltige ternäre Phasen (H-Phase)," *Mh. Chem.*, vol. 94, p. 672, 1963.

[5]   W. Jeitschko, H. Nowotny and F. Benesovsky, "Kohlenstoffhaltige ternäre Phasen (Nb3Al2C und Ta3Al2C)," *Mh. Chem.*, vol. 94, p. 332, 1963.

[6]   W. Jeitschko, H. Nowotny and F. Benesovsky, "Carbides of Formula T2MC. W. Jeitschko," *J. Less-Common Met.*, vol. 7, p. 133, 1964.

[7]   H. Wolfsgruber, H. Nowotny and F. Benesovsky, "Die Kristallstruktur von Ti3GeC2," *Monatsch. Chem.*, vol. 98, p. 2403, 1967.

[8]   W. Jeitschko and H. Nowotny, "Die Kristallstruktur von Ti3SiC2 — ein neuer Komplexcarbid-Typ.," *Monatsch. Chem.*, vol. 98, p. 329, 1967.

[9]   V. H. Nowotny, "Strukturchemie einiger Verbindungen der Übergangsmetalle mit den elementen C, Si, Ge, Sn," *Prog. Solid State Chem.*, vol. 5, p. 27, 1971.

[10]  G. Hägg, "Gesetzmässigkeiten im kristallbau bei hydriden, boriden, carbiden und nitriden der übergangselemente," *Z. Phys. Chem.*, vol. 12, p. 33, 1931.

[11]  M. W. Barsoum and T. El-Raghy, "Synthesis and characterization of a remarkable ceramic: Ti3SiC2," *J. Am. Ceram. Soc.*, vol. 79, p. 1953, 1996.

[12]  M. W. Barsoum, "The MN+1AXN phases: a new class of solids: thermodynamically stable nanolaminates," *Prog. Solid State Chem.*, vol. 28, p. 201, 2000.

[13]  A. L. Ivanovsky, D. L. Novikov and G. P. Shveikin, "Electronic Structure of Ti3SiC2," *Mendeleev Communications*, vol. 3, p. 90, 1995.

[14]  P. Eklund, M. Beckers, U. Jansson and L. Hultman, "The MAX phases: Materials science and thin-film processing," *Thin Solid Films*, vol. 518, p. 1851, 2010.

[15]  N. Lane, S. C. Vogel and M. W. Barsoum, "High-temperature neutron diffraction and the temperature-dependent crystal structures of the MAX phases

Ti3SiC2 and Ti3GeC2," *Phys. Rev. B*, vol. 82, p. 174109, 2010.

[16] M. Magnuson, M. Mattesini, N. Van Nong, P. Eklund and L. Hultman, "Electronic-structure origin of the anisotropic thermopower of nanolaminated Ti3SiC2 determined by polarized x-ray spectroscopy and Seebeck measurements," *Phys. Rev. B*, vol. 85, p. 195134, 2012.

[17] M. Mattesini and M. Magnuson, "Electronic correlation effects in the Cr2GeC Mn+1AXn phase," *J. Phys. Cond. Mat.*, vol. 25, p. 035601, 2013.

[18] M. Magnuson, N. Wassdahl, A. Nilsson, A. Föhlisch, J. Nordgren and N. Mårtensson, "Resonant Auger spectroscopy at the L2,3 shake-up thresholds as a probe of electron correlation effects in nickel," *Phys. Rev. B*, vol. 58, p. 3677, 1998.

[19] P. O. Nilsson, J. Kanski, J. V. Thordson, T. G. Andersson, J. Nordgren, J.-H. Guo and M. Magnuson, "Electronic structure of buried Si layers in GaAs(001) as studied by soft X-ray emission," *Phys. Rev B*, vol. 52, p. R8643, 1995.

[20] M. Magnuson, T. Schmitt, V. N. Strocov, J. Schlappa, A. S. Kalabukhov and L. Duda, " Structure and bonding in amorphous iron carbide thin films Self-doping processes between planes and chains in the metal-to-superconductor transition of YBa2Cu3O6.9," *Scientific Reports*, vol. 4, p. 7017, 2014.

[21] M. Magnuson and C. F. Hague, "Determination of the refractive index at soft X-ray resonances," *J. Elec. Spectrosc. Relat. Phen.*, Vols. 137-140, p. 519–522, 2004.

[22] M. Magnuson, E. Lewin, L. Hultman and U. Jansson, "Electronic structure and chemical bonding of nc-TiC/a-C nanocomposites," *Phys. Rev. B*, vol. 80, p. 235108, 2009.

[23] M. Magnuson, J.-P. Palmquist, M. Mattesini, S. Li, R. Ahuja, O. Eriksson, J. Emmerlich, O. Wilhelmsson, P. Eklund, L. Högberg, L. Hultman and U. Jansson, "Electronic structure investigation of Ti3AlC2, Ti3SiC2, and Ti3GeC2 by soft-X-ray emission spectroscopy," *Phys. Rev. B*, vol. 72, p. 245101, 2005.

[24] M. Magnuson, M. Mattesini, S. Li, C. Höglund, M. Beckers, L. Hultman and O. Eriksson, "Bonding mechanism in the nitrides Ti2AlN and TiN: An experimental and theoretical investigation," *Phys. Rev. B.*, vol. 76, p. 195127, 2007.

[25] M. Magnuson, M. Mattesini, M. Bugnet and P. Eklund, "The origin of anisotropy and high density of states in the electronic structure of Cr2GeC by means of polarized soft X-ray spectroscopy and ab initio calculations," *J. Phys. Cond. Mat.*, vol. 27, p. 415501, 2015.

[26] M. Magnuson, O. Wilhelmsson, M. Mattesini, S. Li, R. Ahuja, O. Eriksson, H. Högberg, L. Hultman and U. Jansson, "Anisotropy in the electronic structure of V2GeC investigated by soft x-ray emission spectroscopy and first-principles theory," *Phys. Rev. B*, vol. 78, p. 035117, 2008.

[27] M. Magnuson, O. Wilhelmsson, J.-P. Palmquist, U. Jansson, M. Mattesini, S. Li, R. Ahuja and O. Eriksson, "Electronic structure and chemical bonding in Ti2AlC investigated by soft x-ray emission spectroscopy," *Phys. Rev. B.*, vol. 74, p. 195108, 2006.

[28] M. F. Cover, O. Warschkow, M. M. M. Bilek and D. R. McKenzie, "A comprehensive survey of M2AX phase elastic properties," *J. Phys. Cond. Mat.*, vol. 21, p. 305403 , 2009.

[29] M. Dahlqvist, B. Alling and J. Rosén, "Stability trends of MAX phases from first principles," *Phys. Rev. B,* vol. 81, p. 220102(R), 2010.

[30] M. Ashton, R. G. Hennig, S. R. Broderick, K. Rajan and S. B. Sinnott, "Computational discovery of stable M2AX phases," *Phys. Rev. B,* vol. 94, p. 054116, 2016.

[31] R. Stumm von Bordwehr, "A History of X-ray absorption fine structure," *Ann. Phys. Fr. ,* vol. 14, no. 4, pp. 377 - 465, 1989.

[32] J. Stöhr, NEXAFS Spectroscopy, Berlin: Springer-Verlag, 1992.

[33] G. Bunker, Introduction to XAFS: A Practical Guide to X-ray Absorption Fine Structure Spectroscopy, Cambridge University Press, 2010.

[34] W. Olovsson, B. Alling and M. Magnuson, "Structure and Bonding in Amorphous Cr1-xCx Nanocomposite Thin Films: X-ray Absorption Spectra and First-Principles Calculations," *J. Phys. Chem. C.,* vol. 120, p. 12890, 2016.

[35] N. A. Rowland, "Preliminary notice of the results accomplished in the manufacture and theory of gratings for optical purposes," *Phil. Mag.,* vol. 5, no. 13, p. 469, 1882.

[36] M. Siegbahn, The spectroscopy of X-rays, London: Milford, 1925.

[37] H. W. B. Skinner, "The Soft X-Ray Spectroscopy of Solids. I. K- and L-Emission Spectra from Elements of the First Two Groups," *Phil. Trans. R. Soc. London Ser. A,* vol. 239, p. 95, 1940.

[38] F. de Groot and A. Kotani, Core Level Spectroscopy of Solids, Boca Raton: CRC Press, 2008.

[39] M. Magnuson, Electronic structure studies using resonant X-ray and photoemission spectroscopy, http://liu.diva-portal.org/smash/get/diva2:319606/FULLTEXT01.pdf, Ed., Uppsala: Acta Universitatis Upsaliensis, Uppsala, Eklundshofs grafiska, 1999. Comprehensive summaries of Uppsala dissertations from the Faculty of Science and Technology, ISSN 1104-232X; 452, ISBN: 91-554-4463-6., 1999.

[40] H. A. Kramers and W. Heisenberg, "Über die Streuung von Strahlung durch Atome," *Zeitschrift für Physik,* vol. 31, no. 1, p. 681–708 , 1925.

[41] D. Harris and M. Bertolucci, Symmetry and Spectroscopy, Oxford University Press, 1978.

[42] V. R. Galakhov, V. A. Trofimova, D. G. Kellerman, Y. N. Blinovskov and V. A. Perelyaev, "X-ray emission and photoelectronspectra of Ti3SiC2," *Solid State Chemistry and Novel Materials Conference,* vol. 1, p. 57, 1996.

[43] N. Medvedeva, D. Novikov, A. Ivanovsky, M. Kuznetsov and A. Freeman, "Electronic properties of Ti3SiC2-based solid solutions," *Phys. Rev. B,* vol. 58, p. 16042, 1998.

[44] M. Magnuson, N. Wassdahl and J. Nordgren, " Energy dependence of Cu L2,3 satellite structures using synchrotron excited X-ray emission spectroscopy," *Phys. Rev. B,* vol. 56, p. 12238, 1997.

[45] P. Blaha, K. Schwarz, G. K. H. Madsen, D. Kvasnicka and J. Luitz, WIEN2k, An Augmented Plane Wave+Local Orbitals Program for Calculating Crystal Properties, Karlheinz Schwarz, Techn. Universität Wien, Austria , 2001.

[46] M. Magnuson, "MAX-Phases Investigated by Soft X-Ray Emission Spectroscopy," *Mechanical Properties and Performance of Engineering*

*Ceramics II: Ceramic Engineering and Science Proceedings,* vol. 27, no. 2, p. 325, 2006.

[47] R. Laskowski and P. Blaha, "Understanding the L2,3 x-ray absorption spectra of early 3d transition elements," *Phys. Rev. B,* vol. 82, p. 205104 , 2010.

[48] D. Coster and R. D. L. Kronig, "New type of Auger effect and its influence on the x-ray spectrum," *Physica,* vol. 2, no. 1-12, p. 13–24, 1935.

[49] A. Furlan, U. Jansson, J. Lu, L. Hultman and M. Magnuson, "Structure and bonding in amorphous iron carbide thin films," *J. Phys. Cond. Mat.,* vol. 27, p. 045002, 2015.

[50] M. M. Magnuson, L.-C. Duda, S. M. Butorin, P. Kuiper and J. Nordgren, "Large magnetic circular dichroism in resonant inelastic x-ray scattering at the Mn L-edge of Mn-Zn ferrite," *Phys. Rev. B,* vol. 74, p. 172409, 2006.

[51] R. Laskowski and N. E. Christensen, "Ab initio calculations of excitons in AlN and Elliott's model," *Phys. Rev. B,* vol. 74, p. 075203, 2006.

[52] Y. Zhou, Z. Sun, X. Wang and S. Chen, "Ab initio geometry optimization and ground state properties of layered ternary carbides Ti3MC2 (M = Al, Si and Ge)," *J. Phys.: Condens. Matter,* vol. 13, p. 10001, 2001.

[53] Y. L. Du, Z. M. Sun, H. Hashimoto and W. B. Tian, "First-principles study of polymorphism in Ta4AlC3," *Solid State Commun.,* vol. 145, p. 461, 2008.

[54] G. Hug and E. Fries, "Full-potential electronic structure of Ti2AlC and Ti2AlN," *Phys. Rev. B,* vol. 65, p. 113104, 2002.

[55] Z. Sun, R. Ahuja, S. Li and J. M. Schneider, "Structure and bulk modulus of M2AIC (M=Ti, V, and Cr)," *Appl. Phys. Lett. 83, 899 (2003).,* vol. 83, p. 899, 2003.

[56] Z. M. Sun and Y. C. Zhou, "Ab initio calculation of titanium silicon carbide," *Phys. Rev. B,* vol. 60, p. 1441, 1999.

[57] Y. Mo, P. Rulis and W. Y. Ching, "Electronic structure and optical conductivities of 20 MAX-phase compounds," *Phys. Rev. B,* vol. 86, p. 165122, 2012.

[58] M. T. Nasir and A. K. M. A. Islam, "MAX phases Nb2AC (A = S, Sn): An ab initio study," *Comp. Mat. Sci.,* vol. 65, pp. 365-371, 2012.

[59] W. Sun, W. Luo and R. Ahuja, "Role of correlation and relativistic effects in MAX phases," *J. Mater. Sci.,* vol. 47, p. 7615, 2012.

[60] M. Ramzan, S. Lebegue and R. Ahuja, "Hybrid exchange-correlation functional study of the structural, electronic, and mechanical properties of the MAX phases," *Appl. Phys. Lett.,* vol. 98, p. 021902, 2011.

[61] R. Ahuja, O. Eriksson, J. M. Wills and B. Johansson, "Electronic structure of Ti3SiC2," *Appl. Phys. Lett.,* vol. 76, p. 2226, 2000.

[62] M. Bugnet, M. Jaouen, V. Mauchamp, T. Cabioc'h and G. Hug, "Experimental and first-principles investigation of the electronic structure anisotropy of Cr2AlC," *Phys. Rev. B,* vol. 90, p. 195116, 2014.

[63] E. H. Kisia, J. A. A. Crossley, S. Myhra and M. W. Barsoum, "Structure and Crystal Chemistry of Ti3SiC2," *Journal of Physics and Chemistry of Solids,* vol. 59, p. 1437–1443, 1998.

[64] S. E. Stoltz, H. I. Starnberg and M. W. Barsoum, "Core level and valence band studies of layered Ti3SiC2 by high resolution photoelectron spectroscopy," *J.*

*Phys. Chem. Solids,* vol. 64, p. 2321, 2003.

[65] M. Magnuson, M. Mattesini, O. Wilhelmsson, J. Emmerlich, J.-P. Palmquist, S. Li, R. Ahuja, L. Hultman, O. Eriksson and U. Jansson, "Electronic structure and chemical bonding in Ti4SiC3 investigated by soft x-ray emission spectroscopy and first principle theory," *Phys. Rev. B.,* vol. 74, p. 205102, 2006.

[66] M. Magnuson, M. Mattesini, C. Höglund, J. Birch and L. Hultman, "Electronic structure and chemical bonding anisotropy investigation in wurtzite AlN," *Phys. Rev. B,* vol. 80, p. 155105, 2009.

[67] M. Magnuson, "Electronic structure investigation of MAX-phases by soft x-ray emission spectroscopy," *MRS Proceedings,* vol. 1023, pp. 1023-JJ09-01, 2007.

[68] M. Magnuson, "Investigation of Ti2AlC and TiC by soft x-ray emission spectroscopy," *Journal of Physics; Conference Series,* vol. 61, no. 1, p. 760, 2007.

[69] G. Hug, M. Jaouen and M. W. Barsoum, "X-ray absorption spectroscopy, EELS, and full-potential augmented plane wave study of the electronic structure of Ti2AlC, Ti2AlN, Nb2AlC, and (Ti0.5Nb0.5)2AlC," *Phys. Rev. B,* vol. 71, p. 24105, 2005.

[70] V. Mauchmap, G. Hug, M. Bugnet, T. Cabioc'h and M. Jaouen, "Anisotropy of Ti2AlN dielectric response investigated by ab initio calculations and electron energy-loss spectroscopy," *Phys. Rev. B,* vol. 81, p. 035109, 2010.

[71] U. von Barth and G. Grossmann, "Dynamical effects in x-ray spectra and the final-state rule," *Phys. Rev. B,* vol. 25, p. 5150, 1982.

[72] M. Magnuson, M. Mattesini, C. Höglund, J. Birch and L. Hultman, "Electronic structure of GaN and Ga investigated by soft x-ray spectroscopy and first-principles methods," *Phys. Rev. B,* vol. 81, p. 085125, 2010.

[73] M. Ramzan, S. Lebégue and R. Ahuja, "Electronic and mechanical properties of Cr2GeC with hybrid functional and correlation effects," *Solid State Commun.,* vol. 152, no. 13, pp. 1147-1149, 2012.

[74] G. Hug, "Electronic structures of and composition gaps among the ternary carbides Ti2MC," *Phys. Rev. B,* vol. 74, p. 184113, 2006.

[75] Y. L. Du, Z. M. Sun and H. Hashimoto, "First-principles study on phase stability and compression behavior of Ti2SC and Ti2AlC," *Physica B,* vol. 405, p. 720, 2010.

[76] M. W. Barsoum, A. Murugaiah, S. R. Kalidindi and T. Zhen, "Kinking Nonlinear Elastic Solids, Nanoindentations, and Geology," *Phys. Rev. Lett.,* vol. 92, p. 255508, 2004.

[77] M. W. Barsoum, L. Farber and T. El-Raghy, "Dislocations, kink bands, and room-temperature plasticity of Ti3SiC2," *Metall. Mater. Trans., A 34, 1727 (1999),* vol. 30, p. 1727, 1999.

[78] C. M. Fang, R. Ahuja, O. Eriksson, S. Li, U. Jansson, O. Wilhelmsson and L. Hultman, "General trend of the mechanical properties of the ternary carbides M3SiC2 (M=transition metal)," *Phys. Rev. B,* vol. 74, p. 054106, 2006.

[79] H. Z. Zhang and S. Q. Wang, "First-principles study of Ti3AC2 (A = Si, Al) (001) surfaces," *Acta Mater.,* vol. 55, p. 4645, 2007.

[80] N. I. Medvedeva and A. J. Freeman, "Cleavage fracture in Ti3SiC2 from first-principles," *Scripta Mater.,* vol. 58, p. 671, 2008.

[81]  I. R. Shein, I. S. Kiiko, Y. N. Makurin, M. A. Gorbunova and A. L. Ivanovskiĭ, "Elastic parameters of single-crystal and polycrystalline wurtzite-like oxides BeO and ZnO: Ab initio calculations," *Physics of the Solid State*, vol. 49, p. 1015, 2007.

[82]  V. Mauchamp, W. Yu, L. Gence, L. Piraux, C. T., V. Gauthier, P. Eklund and S. Dubois, "Anisotropy of the resistivity and charge-carrier sign in nanolaminated Ti2AlC: Experiment and ab initio calculations," *Phys. Rev. B*, vol. 87, p. 235105, 2013.

[83]  O. Wilhelmsson, M. Råsander, M. Carlsson, B. Sanyal, U. Wiklund, O. Eriksson and U. Jansson, "Design of Nanocomposite Low-Friction Coatings," *Adv. Funct. Mater.*, vol. 17, no. 10, p. 1611, 2007.

[84]  A. Grechnev, R. Ahuja and O. Eriksson, "Balanced crystal orbital overlap population—a tool for analysing chemical bonds in solids," *J. Phys.: Condens. Matter*, vol. 15, p. 7751, 2003.

[85]  M. Barsoum, T. H. Scabarozi, S. Amini, H. J. D. and S. E. Lofland, "Electrical and Thermal Properties of Cr2GeC," *J. Am. Ceram. Soc.*, vol. 94, p. 4123, 2011.

[86]  M. K. Drulis, H. Drulis, A. E. Hackemer, O. Leaffer, J. Spanier, S. Amini, M. W. Barsoum, T. Guilbert and E.-R. T., "On the heat capacities of Ta 2 AlC, Ti 2 SC, and Cr 2 GeC," *J. Appl. Phys.*, vol. 104, p. 023526, 2008.

[87]  H.-I. Yoo, M. W. Barsoum and T. El-Raghy, "Materials science: Ti3SiC2 has negligible thermopower," *Nature*, vol. 407, p. 581, 2000.

[88]  L. Chaput, G. Hug, P. Pecheur and H. Scherrer, "Anisotropy and thermopower in Ti3SiC2," *Phys. Rev. B*, vol. 71, p. 121104, 2005.

[89]  E. H. Kisi, J. F. Zhang, O. Kirstein, D. P. Riley, M. J. Styles and A. M. Paradowska, "Shear stiffness in nanolaminar Ti3SiC2 challenges ab initio calculations," *J. Phys. Condens. Mat.*, vol. 22, p. 162202, 2010.

[90]  A. Mendoza-Galván, M. Rybka, K. Järrendahl, H. Arwin, M. Magnuson, L. Hultman and M. Barsoum, "Spectroscopic ellipsometry study on the dielectric function of bulk Ti2AlN, Ti2AlC, Nb2AlC, NbTiAlC, and Ti3GeC2 MAX phases," *J. Appl. Phys.*, vol. 109, p. 013530, 2011.

[91]  J. E. Spanier, S. Gupta, M. Amer and M. W. Barsoum, "Vibrational behavior of the Mn+1AXn phases from first-order Raman scattering (M=Ti, V, Cr, A=Si, X(C, N)," *Phys. Rev. B*, vol. 71, p. 012103, 2005.

[92]  M. Amer, M. W. Barsoum, E.-R. T., I. Weiss, S. Leclair and D. Liptak, "The Raman spectrum of Ti3SiC2," *J. Appl. Phys.*, vol. 84, p. 5817, 1998.

[93]  S. Baroni, S. de Gironcoli, A. Dal Corso and P. Giannozzi, "Phonons and related crystal properties from density-functional perturbation theory," *Rev. Mod. Phys.*, vol. 73, p. 515, 2001.

[94]  "Quantum-ESPRESSO," [Online]. Available: http://www.quantum-espresso.org and http://www.pwscf.org. .

[95]  E. Isaev, "QHA project," [Online]. Available: http://qe-forge.org/qha.

[96]  A. Togo, L. Chaput, I. Tanaka and G. Hug, "First-principles phonon calculations of thermal expansion in Ti3SiC2, Ti3AlC2, and Ti3GeC2," *Phys. Rev. B*, vol. 81, p. 174301, 2010.

[97]  J. E. Spanier, S. Gupta, M. Amer and M. W. Barsoum, "First-order Raman scattering from the Mn+1AXn phases," *Phys. Rev. B*, vol. 71, p. 12103, 2005.

[98]  J. Wang, Y. Zhou, Z. Lin, F. Meng and F. Li, "Raman active modes and heat capacities of Ti2AlC and Cr2AlC ceramics: first principles and experimental investigations," *Appl. Phys. Lett.*, vol. 86, p. 101902, 2005.

[99]  O. D. Leaffer, S. Gupta, M. W. Barsoum and J. E. Spanier, "On the Raman scattering from selected M2AC compounds," *J. Mater. Res.*, vol. 22, p. 2651, 2007.

[100] F. Mercier, O. Chaix-Pluchery, T. Ouisse and D. Chaussende, "Raman scattering from Ti3SiC2 single crystals," *Appl. Phys. Lett.*, vol. 98, p. 081912, 2011.

[101] N. J. Lane, M. Naguib, V. Presser, H. G., L. Hultman and M. W. Barsoum, "First-order Raman scattering of the MAX phases Ta4AlC3, Nb4AlC3, Ti4AlN3 and Ta2AlC," *J. Raman Spectrosc.*, vol. 43, pp. 954-958, 2012.

[102] V. Presser, M. Naguib, L. Chaput, A. Togo, G. Hug and M. W. Barsoum, "First-order Raman scattering of the MAX phases: Ti2AlN, Ti2AlC0.5N0.5, Ti2AlC, (Ti0.5V0.5)2AlC, V2AlC, Ti3AlC2 and Ti3GeC2," *J. Raman Spectrosc.*, vol. 43, pp. 168-172, 2012.

[103] L. E. Toth, Transition Metal Carbides and Nitrides, New York: Academic Press, 1971.

[104] B. Manoun, H. Yang, S. K. Saxena, A. Ganguly, M. W. Barsoum, Z. X. Liu, M. Lachkar and B. El-Bali, "Infrared spectrum and compressibility of Ti3GeC2 to 51 GPa," *J. Alloys Compd.*, vol. 433, p. 265, 2007.

[105] S. Li, R. Ahuja, M. W. Barsoum, P. Jena and B. Johansson, "Optical properties of Ti3SiC2 and Ti4AlN3," *Appl. Phys. Lett.*, vol. 92, p. 221907, 2015.

[106] N. Haddad, E. Garcia-Caurel, L. Hultman, M. W. Barsoum and G. Hug, "Dielectric properties of Ti2AlC and Ti2AlN MAX phases: The conductivity anisotropy," *J. Appl. Phys.*, vol. 104, p. 023531, 2008.

[107] G. Hug, P. Eklund and A. Orchowski, "Orientation dependence of electron energy loss spectra and dielectric functions of Ti3SiC2 and Ti3AlC2," *Ultramicroscopy,* vol. 110, no. 8, p. 1054–1058, 2010.

[108] X. He, Y. Bai, Y. Chen, C. Zhu, M. Li and M. W. Barsoum, "Phase Stability, Electronic Structure, Compressibility, Elastic and Optical Properties of a Newly Discovered Ti3SnC2: A First-Principle Study," *J. Am. Ceram. Soc.*, vol. 94, pp. 3907-3914, 2011.

[109] M. B. Kanoun, S. Goumri-Said and A. H. Reshak, "Theoretical study of mechanical, electronic, chemical bonding and optical properties of Ti2SnC, Zr2SnC, Hf2SnC and Nb2SnC," *Comp. Mat. Science* , vol. 47, no. 2, p. 491–500, 2009.

[110] O. Kyriienko and I. A. Shelykh, "Angle-resolved reflectance and surface plasmonics of the MAX phases," *Optics Letters,* vol. 36, no. 20, pp. 3966-3968, 2011.

[111] C. Li, B. Wang, Y. Li and R. Wang, "First-principles study of electronic structure, mechanical and optical properties of V4AlC3," *J. Phys. D: Appl. Phys.*, vol. 42, p. 065407, 2009.

[112] M. Mattesini, R. Ahuja and B. Johansson, "Cubic Hf3N4 and Zr3N4: A class of hard materials," *Phys. Rev. B*, vol. 68, p. 184108, 2003.

[113] M. W. Barsoum and M. Radovic, "Elastic and Mechanical Properties of the

MAX Phases," *Annual Review of Materials Research,* vol. 41, pp. 195-227, 2011.

[114] "MedeA-VASP, Material Designs Inc.," [Online]. Available: http://www.materialsdesign.com.

[115] Y. L. Du, Z.-M. Sun, H. Hashimoto and M. W. Barsoum, "Electron correlation effects in the MAX phase Cr2GeC from first principles," *J. Appl. Phys.,* vol. 109, p. 063707, 2011.

[116] M. Shamma, V. Presser, B. Clausen, D. Brown, O. Yeheskel, S. Amini and M. W. Barsoum, "On the response of titanium sulfocarbide to stress studied by in situ neutron diffraction and the elastoplastic self-consistent approach," *Scripta Materialia,* vol. 65, p. 573–576, 2011.

[117] B. Clausen, T. Lorentzen, M. A. M. Bourke and M. R. Daymond, "Lattice strain evolution during uniaxial tensile loading of stainless steel," *Materials Science and Engineering: A ,* vol. 259, pp. 17-24, 1999.

[118] B. Clausen, T. Lorentzen and T. Leffers, "Self-consistent modelling of the plastic deformation of f.c.c. polycrystals and its implications for diffraction measurements of internal stresses," *Acta Materialia,* vol. 46, p. 3087–3098, 1998.

[119] B. Clausen, C. N. Tomé, D. W. Brown and S. R. Agnew, "Reorientation and stress relaxation due to twinning: Modeling and experimental characterization for Mg," *Acta Materialia,* vol. 56, p. 2456–2468, 2008.

[120] P. A. Turner and C. N. Tomé, "A study of residual stresses in Zircaloy-2 with rod texture," *Acta Metallurgica et Materialia,* vol. 42, pp. 4143-4153, 1994.

[121] M. Born and K. Huang, Dynamical Theory of Crystal Lattices, Clarendon, Oxford, 1956.

[122] R. Hill, "The Elastic Behaviour of a Crystalline Aggregate," *Proceedings of the Physical Society. Section A,* vol. 65, p. 349, 1952.

[123] Y. L. Du, Z. M. Sun, H. Hashimoto and M. W. Barsoum, "Theoretical investigations on the elastic and thermodynamic properties of Ti2AlC0.5N0.5 solid solution," *Physics Letters A,* vol. 374, p. 78–82, 2009.

[124] M. B. Kanoun, S. Goumri-Said and M. Jaouen, "Steric effect on the M site of nanolaminate compounds M2SnC (M = Ti,Zr, Hf and Nb)," *J. Phys.: Cond. Mat.,* vol. 21, p. 045404, 2009.

[125] D. Music, Z. Sun, R. Ahuja and J. M. Schneider, "Coupling in nanolaminated ternary carbides studied by theoretical means: The influence of electronic potential approximations," *Phys. Rev. B,* vol. 73, p. 134117, 2006.

[126] M. Mattesini, M. Magnuson, F. Tasnádi, C. Höglund, I. A. Abrikosov and L. Hultman, "Elastic properties and electrostructural correlations in ternary scandium-based cubic inverse perovskites: A first-principles study," *Phys. Rev. B,* vol. 79, p. 125122, 2009.

[127] W. Tian, P. Wang, G. Zhang, Y. Kan and Y. Li, "Mechanical Properties of Cr2AlC Ceramics," *J. Am. Ceram. Soc.,* vol. 90, p. 1663, 2007.

[128] S. F. Pugh, "XCII. Relations between the elastic moduli and the plastic properties of polycrystalline pure metals," *Philos. Mag.,* vol. 45, p. 823, 1954.

[129] O. L. Anderson, Equation of States of Solids for Geophysics and Ceramic Science, New York: Oxford University Press, 1995.

[130] A. M. Ibrahim, *Nucl. Instrum. Methods Phys. Res. B*, vol. 34, p. 135, 1988.

[131] M. K. Drulis, A. Czopnik, H. Drulis and M. W. Barsoum, "Low temperature heat capacity and magnetic susceptibility of Ti3SiC2," *J. Appl. Phys.*, vol. 95, pp. 128-133, 2004.

[132] P. Finkel, M. W. Barsoum and T. El-Raghy, "Low temperature dependencies of the elastic properties of Ti4AlN3, Ti3Al1.1C1.8, and Ti3SiC2," *J. Appl. Phys.*, vol. 87, pp. 1701-1703, 2000.

[133] R. Pompuch, J. Lis, L. Stobievski and M. Tymkiewicz, "Effect of boron, nitrogen, and oxygen impurities on the electronic structure and deformation behavior of Ti3SiC2," *J. Eur. Ceram. Soc.*, vol. 5, p. 283, 1989.

[134] J. Lis, R. Pampuch, J. Piekarczyk and L. Stobierski, "New ceramics based on Ti3SiC2," *Ceram. Int.*, vol. 19, p. 219, 1993.

[135] S. Amini, A. Zhou, S. Gupta, A. DeVillier, P. Finkel and M. W. Barsoum, "Synthesis and elastic and mechanical properties of Cr2GeC," *J. Mater. Res.*, vol. 23, pp. 2157-2165 , 2008.

[136] A. Bouhemadou, "Calculated structural, electronic and elastic properties of M2GeC (M=Ti, V, Cr, Zr, Nb, Mo, Hf, Ta and W)," *Appl. Phys. A,* vol. 96, p. 959., 2009.

[137] J. Emmerlich, D. Music, A. Houben, R. Dronskowski and J. M. Schneider, "Systematic study on the pressure dependence of M2AlC phases (M=Ti,V,Cr,Zr,Nb,Mo,Hf,Ta,W)," *Phys. Rev. B*, vol. 76, p. 224111, 2007.

[138] Z. Sun, S. Li, R. Ahuja and J. M. Schneider, "Calculated elastic properties of M2AlC (M=Ti, V, Cr, Nb and Ta)," *Solid State Commun.*, vol. 129, p. 589–592, 2004.

[139] J. Wang and Y. Zhou, "Dependence of elastic stiffness on electronic band structure of nanolaminate M2AlC (M=Ti,V,Nb, and Cr) ceramics," *Phys. Rev. B*, vol. 69, p. 214111, 2004.

[140] V. Tvergaard and J. W. Hutchinson, "Microcracking in ceramics induced by thermal expansion or elastic anisotropy," *J. Am. Ceram. Soc.*, vol. 71, p. 157, 1988.

[141] J. M. Schneider, Z. M. Sun, R. Mertens, F. Uestel and R. Ahuja, "Ab initio calculations and experimental determination of the structure of Cr2AlC," *Solid State Commun.*, vol. 130, no. 7, p. 445–449, 2004.

[142] W. Luo and R. Ahuja, "Magnetic Fen+1ACn (n = 1, 2, 3,and A = Al, Si, Ge) phases: from ab initio theory," *J. Phys.: Condens. Matter*, vol. 20, p. 064217 , 2008.

[143] M. Magnuson, M. Mattesini, C. Höglund, I. A. Abrikosov, J. Birch and L. Hultman, "Electronic structure investigation of the cubic inverse perovskite Sc3AlN," *Phys. Rev. B,* vol. 78, p. 235102, 2008.

[144] M. Dahlqvist, B. Alling, I. A. Abrikosov and J. Rosen, "Magnetic nanoscale laminates with tunable exchange coupling from first principles," *Phys. Rev. B,* vol. 84, p. 220403(R), 2011.

[145] S. Lin, P. Tong, B. S. Wang, Y. N. Huang, W. J. Lu, D. F. Shao, B. C. Zhao, W. H. Song and Y. P. Sun, "Magnetic and electrical/thermal transport properties of Mn-doped Mn+1AXn phase compounds Cr2−x Mn x GaC (0 ≤ x ≤ 1)," *J. Appl. Phys.*, vol. 113, p. 053502, 2013.

[146] A. S. Ingason, A. Mockute, M. Dahlqvist, F. Magnus, S. Olafsson, U. B. Arnalds, B. Alling, I. A. Abrikosov, B. Hjörvarsson, P. O. Å. Persson and J. Rosen, "Magnetic Self-Organized Atomic Laminate from First Principles and Thin Film Synthesis," *Phys. Rev. Lett.*, vol. 110, p. 195502, 2013.

[147] M. Jaouen, M. Bugnet, N. Jaouen, P. Ohresser, V. Mauchamp, T. Cabioc'h and A. Rogalev, "Experimental evidence of Cr magnetic moments at low temperature in Cr2A(A=Al, Ge)C," *J. Phys. Cond. Mat.*, vol. 26, no. 17, p. 176002, 2014.

[148] W. Zhou, L. Liu and P. Wu, "First-principles study of structural, thermodynamic, elastic, and magnetic properties of Cr2GeC under pressure and temperature," *J. Appl. Phys.*, vol. 106, no. 3, pp. 033501-033501-7, 2009.

[149] J. Stöhr and H. C. Siegmann, Magnetism From Fundamentals to Nanoscale Dynamics, Berlin: Springer Verlag, 2006.

[150] J. Heyd, G. E. Scuseria and M. Ernzerhof, "Hybrid functionals based on a screened Coulomb potential," *J. Chem. Phys.*, vol. 118, p. 8207, 2003.

[151] M. Dahlqvist, B. Alling and J. Rosen, "A critical evaluation of GGA+U modeling for atomic, electronic and magnetic structure of Cr2AlC, Cr2GaC and Cr2GeC," *J. Phys.: Condens. Matter*, vol. 27, p. 095601 , 2015.

[152] M. Magnuson, S. M. Butorin, C. Såthe, J. Nordgren and P. Ravindran, "Spin transition in LaCoO3 investigated by resonant soft X-ray emission spectroscopy," *Europhys. Lett.*, vol. 68, p. 289 , 2004.

[153] M. Dahlqvist, B. Alling and J. Rosén, "Correlation between magnetic state and bulk modulus of Cr2AlC," *J. Appl. Phys.*, vol. 113, p. 216103, 2013.

[154] M. Ramzan, S. Lebègue and R. Ahuja, "Correlation effects in the electronic and structural properties of Cr2AlC," *Phys. Status Solidi*, vol. 5, p. 122, 2011.

[155] N. Li, C. C. Dharmawardhana, K. L. Yao and W. Y. Ching, "Theoretical characterization on intrinsic ferrimagnetic phase in nanoscale laminated Cr2GeC," *Solid State Commun.*, vol. 174, p. 43, 2013.

[156] M. Magnuson, "Induced magnetism at the interfaces of a Fe/V superlattice investigated by resonant magnetic x-ray scattering," *Journal of Magnetism and Magnetic Materials*, vol. 422, p. 362–366, 2016.

[157] A. Mockute, M. Dahlqvist, J. Emmerlich, L. Hultman, J. M. Schneider, P. O. Å. Persson and J. Rosen, "Synthesis and ab initio calculations of nanolaminated (Cr,Mn)2AlC compounds," *Phys. Rev. B,* vol. 87, p. 094113, 2013.

[158] M. Mockute, J. Lu, E. J. Moon, M. Yan, B. Anasori, S. J. May, M. W. Barsoum and J. Rosén, "Solid Solubility and Magnetism upon Mn Incorporation in the Bulk Ternary Carbides Cr2AlC and Cr2GaC," *Mat. Res. Lett.*, vol. 3, p. 16, 2014.

[159] A. S. Ingason, M. Dahlqvist and J. Rosén, "Magnetic MAX phases from theory and experiments; a review," *J. Phys. Cond. Mat.*, vol. 28, p. 433003, 2016.

[160] M. Naguib, V. N. Mochalin, M. W. Barsoum and Y. Gogotsi, "25th Anniversary Article: MXenes: A New Family of Two-Dimensional Materials," *Adv. Mater.*, vol. 26, p. 992, 2013.

[161] M. Naguib, M. Kurtoglu, V. Presser, J. Lu, J. Niu, M. Heon, L. Hultman, Y. Gogotsi and M. W. Barsoum, "Two-Dimensional Nanocrystals Produced by Exfoliation of Ti3AlC2," *Adv. Mater.*, vol. 23, p. 4248, 2011.